\documentclass{emulateapj}
\usepackage{apjfonts}
\usepackage{graphicx}
\usepackage{amsmath}
\usepackage{color}
\usepackage{natbib}
\usepackage{threeparttable}
\newcommand{\bim}{\begin{itemize}}
\newcommand{\eim}{\end{itemize}}
\newcommand{\beq}{\begin{equation}}
\newcommand{\eeq}{\end{equation}}

\newcommand{\muk}{$\mu\mathrm{K}$}

\newcommand{\msol}{h^{-1}M_{\Sun}}
\newcommand{\Msol}{M_{\Sun}}
\newcommand{\mpc}{h^{-1}\mathrm{Mpc}}

\newcommand{\dd}{\mathrm{d}}
\newcommand{\planck}{{\sl Planck}}

\begin{document}
\title{Simulations of the Pairwise Kinematic Sunyaev-Zel'dovich Signal}

\author{Samuel Flender$^{1,2}$, Lindsey Bleem$^{1,2}$, Hal Finkel$^{3}$, Salman Habib$^{1,2,4}$, Katrin Heitmann$^{1,2,4}$, Gilbert Holder$^{5}$}
\affil{$^1$HEP Division, Argonne National Laboratory, 9700 S. Cass Ave., Lemont, IL 60439, USA}  
\affil{$^2$Kavli Institute for Cosmological Physics, The University of Chicago, Chicago, IL 60637, USA}
\affil{$^3$ALCF Division, Argonne National Laboratory, Lemont, IL 60439, USA}
\affil{$^4$MCS Division, Argonne National Laboratory, 9700 S. Cass Ave., Lemont IL 60439, USA}
\affil{$^5$Department of Physics, McGill University, 3600 rue University, Montreal, QC, H3A 2T8, Canada}

\begin{abstract}
The pairwise kinematic Sunyaev-Zel'dovich (kSZ) signal from galaxy clusters is a probe of their line-of-sight momenta, and thus a potentially valuable source of cosmological information. In addition to the momenta, the amplitude of the measured signal depends on the properties of the intra-cluster gas and observational limitations such as errors in determining cluster centers and redshifts. 
In this work we simulate the pairwise kSZ signal of clusters at $z<1$, using the output from a cosmological $N$-body simulation and including the properties of the intra-cluster gas via a model that can be varied in post-processing. 
We find that modifications to the gas profile due to star formation and feedback reduce the pairwise kSZ amplitude of clusters by ${\sim50\%}$, relative to the naive ``gas traces mass'' assumption. 
We demonstrate that mis-centering can reduce the overall amplitude of the pairwise kSZ signal by up to 10\%, while redshift errors can lead to an almost complete suppression of the signal at small separations.
We confirm that a high-significance detection is expected from the combination of data from current-generation, high-resolution CMB experiments, such as the South Pole Telescope, and cluster samples from optical photometric surveys, such as the Dark Energy Survey.
Furthermore, we forecast that future experiments such as Advanced ACTPol in conjunction with data from the Dark Energy Spectroscopic Instrument will yield detection significances of at least $20\sigma$, and up to $57\sigma$ in an optimistic scenario. 
Our simulated maps are publicly available at \texttt{http://www.hep.anl.gov/cosmology/ksz.html}.

\end{abstract}


\section{Introduction}

On scales smaller than the homogeneity sphere, clusters of galaxies move towards each other due to their mutual gravitational attraction. A measurement of this pairwise motion has the potential of being a valuable consistency check for the standard cosmological model, as well as placing constraints on extensions to that model, such as dynamical dark energy and modified gravity models.

From the point of view of an observer, a cluster pair would be seen with opposite line-of-sight velocities: the cluster partner that is further away has a velocity component towards the observer, and vice versa. The pairwise velocity of clusters, i.e.,\ the average velocity at which clusters at a given distance move towards each other, can in principle be estimated using only the information about their line-of-sight peculiar velocities \citep{Ferreira:1998id}. In practice however, peculiar velocities are difficult to measure, as they require precise measurement of distances as well as redshifts.

The Sunyaev-Zel'dovich (SZ) effect \citep{Zeldovich:1969ff} is caused by inverse Compton scattering of cosmic microwave background (CMB) photons with free, high-energy electrons, and can be further decomposed into the thermal SZ (tSZ) and the kinematic SZ (kSZ) effects. For clusters of galaxies, the dominating SZ component is the tSZ effect, which is sourced by the electrons that reside inside the hot intra-cluster gas. The tSZ signature can thus be used to detect new clusters in CMB data (see e.g., \citealt{hasselfield13}, \citealt{Bleem:2014iim}, \citealt{planck15XXVII}). The kSZ effect on the other hand is not only sourced by the hot gas inside clusters, but in general by electrons moving with high bulk velocities, including groups, filaments, and the inter-galactic medium. The kSZ signal from clusters is thus a potential probe of their line-of-sight peculiar velocities. Prospects for reconstructing the peculiar velocities of clusters via their kSZ signal have been discussed in numerous works, including\ \citet{Sunyaev:1980nv, 1991ApJ...372...21R,2001AA...374....1A}.

The detection of this signal for single clusters is challenging owing to its small amplitude---which is typically of the order of only a few $\mu$K---and its identical spectral dependence to the CMB (though see \citet{sayer13} for an exceptional system). However, because of the pairwise motion of clusters, the kSZ effect created by a cluster pair creates a distinct pattern in the CMB, consisting of a subtle temperature increment at the location of  clusters moving towards the observer and a decrement at the location of clusters moving away. We call this distinct CMB pattern created by a cluster pair the \emph{pairwise kSZ signal}.

The first detection of the pairwise kSZ signal was reported in \citet{Hand:2012ui}, using high-resolution CMB data from the Atacama Cosmology Telescope (ACT) in conjunction with the Baryon Oscillation Spectroscopic Survey (BOSS) spectroscopic catalog \citep{2012ApJS..203...21A}. The authors report a rejection of the null hypothesis of no kSZ signal with a $p$-value of 0.002. More recently, evidence for the pairwise kSZ signal was also found in CMB data from \planck \ \citep{Ade:2015lza}, using the Central Galaxy Catalog derived from the Sloan Digital Sky Survey \citep{Abazajian:2008wr}. The authors report a statistical significance of $1.8-2.5\sigma$ for a detection of the pairwise kSZ signal. Looking forward, \citet{Keisler:2012eg} showed that a high-significance detection of the pairwise kSZ signal is expected by combining CMB data from the South Pole Telescope (SPT; \citealt{2011PASP..123..568C}) with positions of galaxy clusters obtained from the Dark Energy Survey\footnote{http://www.darkenergysurvey.org} (DES). The authors demonstrate that, even in the presence of redshift errors inherent in a photometric survey, a detection significance of $8-13\sigma$ is possible, depending on the lower mass threshold of the cluster sample. The forecast in \citet{Keisler:2012eg} is based on simulated CMB maps from \citet{Sehgal:2009xv}.

An alternative approach for measuring the kSZ signal is the cross-correlation of the CMB map with a velocity field that has been reconstructed from the positions of galaxies, using linear perturbation theory. Using this velocity-reconstruction method, \citet{Ade:2015lza} report a detection of the kSZ signal with a statistical significance of $3.0-3.7\sigma$. \citet{Schaan:2015uaa} report a detection of the kSZ signal with a statistical significance of $2.9-3.3\sigma$ in CMB data from ACTPol, using the velocity field reconstructed from the CMASS catalog of the BOSS DR10 catalog \citep{2014ApJS..211...17A}. \citet{Li:2014mja} forecast that an even higher significance detection (7.7$\sigma$) of the kSZ signal for \planck \ might be possible, using the velocity field reconstructed with the MaxBCG catalog \citep{2007ApJ...660..221K}.

From an astrophysical point of view, a measurement of the kSZ signal of clusters could be an interesting probe of what is known in the literature as the ``missing baryon problem'' \citep{Bregman:2007ac}; hydrodynamic simulations suggest that most of the baryons inside halos are in a diffuse phase with a temperature range of $10^5-10^7$\,K, which is both too hot to form stars and too cold to be seen in X-ray observations (see \citealt{Cen:2006by} and references therein). These baryons are thus ``missing'' in the sense that they cannot be observed using these methods. However, the baryons in the diffuse phase do contribute to the kSZ signal (see, e.g., \citealt{2008A&A...490...25H, Ho:2009iw, 2011MNRAS.413..628S}). 

From a theoretical point of view, the question arises as to whether---and to what extent---a measurement of the pairwise kSZ signal can be used for constraining cosmological models. For instance, it has been argued that the pairwise velocity inferred from the pairwise kSZ signal can be used to constrain modified-gravity models \citep{Kosowsky:2009nc, Keisler:2012eg, Mueller:2014nsa}, dark energy parameters \citep{Bhattacharya:2007sk, Bhattacharya:2008qc, Mueller:2014nsa}, or the sum of the neutrino masses \citep{Mueller:2014dba}. In addition to the pairwise kSZ measurement, it was also shown that a measurement of the kSZ power spectrum can yield constraints on dark energy models \citep{Ma:2013taq} and modified gravity models \citep{Bianchini:2015iaa}. \citet{Li:2014mja} propose that a measurement of the kSZ signal using the velocity-reconstruction method can function as a test of General Relativity. Other methods for detecting the kSZ signal from clusters and extracting cosmological information from that signal were discussed in \citet{HernandezMonteagudo:2005ys}. All of these studies were performed under differing degrees of observational realism.

The pairwise kSZ signal is not only sensitive to the gravitational attraction of clusters, but also to the physics related to the intra-cluster gas (see, e.g., \citealt{DeDeo:2005yr}). 
\citet{2010ApJ...725...91B} demonstrated that astrophysical effects such as star formation, feedback, and non-thermal pressure, have a significant impact on the amplitude of the tSZ power spectrum (see also \citealt{Shaw:2010mn, 2012ApJ...758...75B}).
The uncertainties in the kSZ amplitude due to the unknown gas physics have been studied in \citet{Shaw:2011sy}. These authors use an analytic model in order to estimate the impact of astrophysical effects on the kSZ power spectrum, implementing a so-called gas window function measured in hydrodynamical simulations in order to account for baryonic physics. The authors find that cooling and star formation can reduce the kSZ power at $\ell=3000$ by up to $\sim30\%$. This result stresses the importance of astrophysical uncertainties when interpreting the amplitude of the measured pairwise kSZ signal in a cosmological context. For recent hydrodynamical simulations including the tSZ and kSZ effects see also \citet{Dolag:2013hj,Dolag:2015dta}. \cite{Dolag:2013hj} showed that the amplitude of the pairwise kSZ signal reported in \cite{Hand:2012ui} is in qualitative agreement with their simulations, although it strongly depends on the mass cut applied to the cluster sample.

The purpose of the present work is to complement previous studies of the pairwise kSZ signal by investigating how the performance of the standard pairwise estimator depends on systematic effects like redshift errors, cluster mis-centering, and scatter in the relation between the cluster mass and richness, along with the effects due to different assumptions regarding the treatment of intra-cluster gas. We also investigate the effect of the filter choice as applied to the CMB maps. In particular, we compare two different filtering strategies: the matched filter, which is the filter with minimum variance for a signal with known profile embedded into noise with a known power spectrum, and the compensated top-hat filter, which makes no assumptions about the signal profile nor about the noise spectrum. While the analysis in \citet{Hand:2012ui} is based on the former, the analysis in \citet{Ade:2015lza} is based on the latter.

In this work we model the SZ effect using the output from a large $N$-body simulation, the \emph{MiraU} simulation, which we will describe in detail below\footnote{Ours is not the first work to study secondary CMB anisotropies using the output from an $N$-body simulation; see \cite{Sehgal:2009xv} for a more detailed discussion.}. In particular, we consider three different models for the kSZ signal. In Model\,I, we assume that baryons trace the dark matter on all scales, which likely over-estimates the true kSZ amplitude. In Model\,II we follow a gas prescription for the intra-cluster gas, based on the gas model from \cite{Shaw:2010mn}, neglecting the component from filaments and the intergalactic medium. Finally, in Model\,III we combine Model\,II with the diffuse kSZ component from Model\,I. Model\,III is thus our most realistic kSZ model. By comparing the pairwise kSZ signal from different models we are able to quantitatively estimate the impact of astrophysical effects on the kSZ signal from clusters. In particular, we find that accounting for star formation and feedback in a way that is consistent with CMB observations, reduces the kSZ amplitude by a factor of $\sim$2 relative to Model\,I.

The outline of the rest of the paper is as follows: In Section~2 we describe the details of the simulation and the derived kSZ and tSZ maps, with a focus on a comparison of the maps and power spectra for different kSZ models. In Section\,3 we introduce the matched filter and the pairwise estimator, and discuss the results for the pairwise kSZ signal obtained from the simulations. Most of the discussion is framed in the context of SPT and DES observations, but we also include forecasts for several next generation experiments. We conclude with a discussion of the results and future prospects in Section\,4.


\section{CMB simulations}

\begin{figure}
\centering
\includegraphics[scale=0.5]{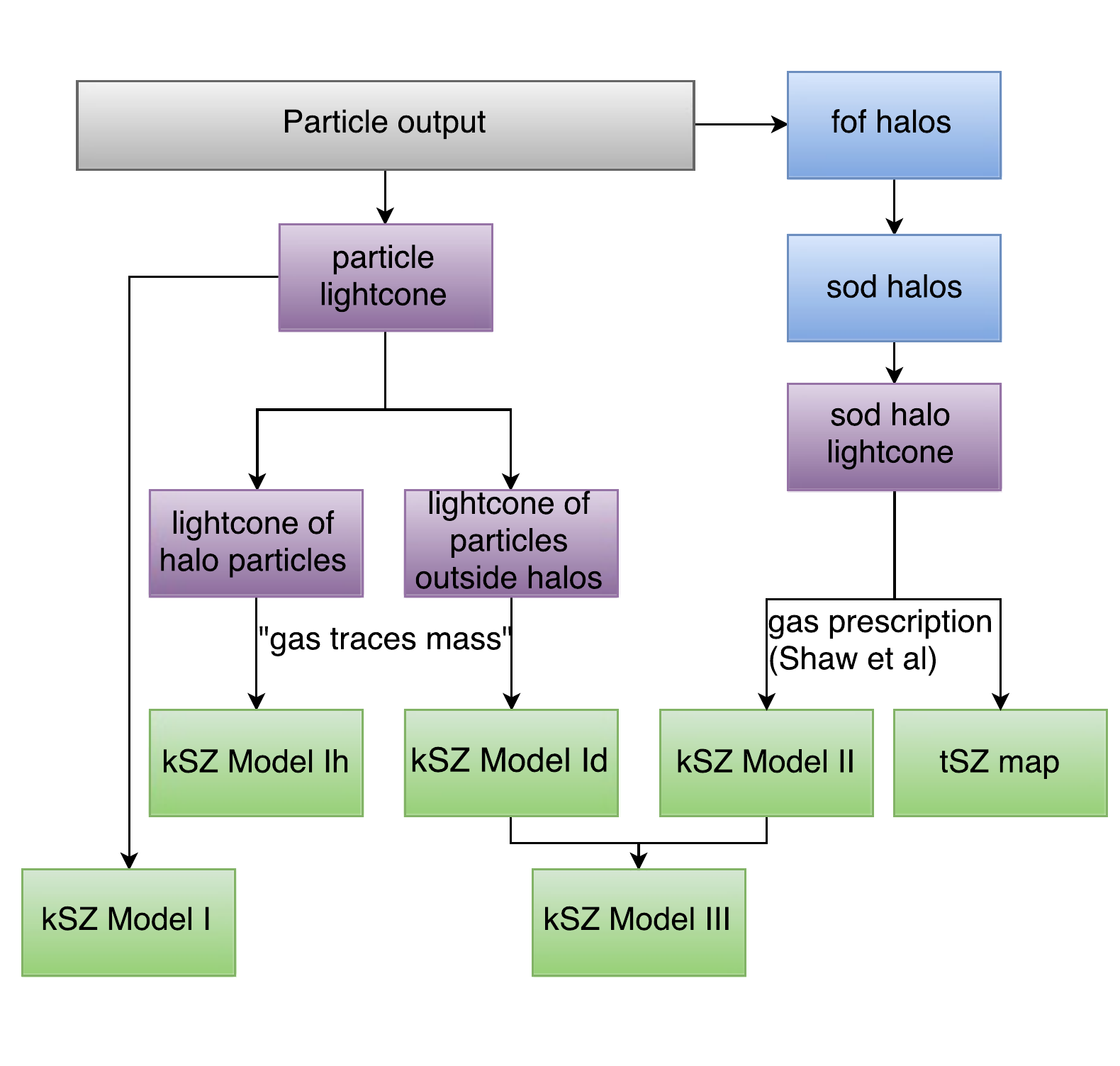}
\caption{\label{fig:flowchart} Flowchart summarizing our post-processing pipeline from the raw particle output of the $N$-body simulation to the kSZ and tSZ maps. The particle lightcone generated from the raw particle data is split into halo-particles and non-halo-particles, leading to two separate kSZ maps, which we call Model\,Ih and Model\,Id, respectively. The sum of these two components is Model\,I. We run an fof-halofinder on the raw particle data, and generate a catalog with spherical overdensity masses (sod halo catalog) by computing $M_{200}$ for each halo with more than 1000 particles. From this halo catalog we generate a lightcone, and from that both the kSZ Model\,II map and the tSZ map, using a gas prescription for the halos which we detail in Section \ref{sec:gasmodel}. Finally, kSZ Model\,III, our most realistic model, is obtained by summing Model\,II and Model\,Id.}
\end{figure}

\label{sec:simulations}
The purpose of this work is to obtain a better understanding of the astrophysical and systematic uncertainties in the context of the pairwise kSZ signal. To this end, we create simulated kSZ and tSZ maps, using the output from an $N$-body simulation. In this section we describe how we generate a full-sky lightcone from the raw particle output from the simulation, and how we generate full-sky kSZ and tSZ maps from that lightcone. We present a flowchart showing all post-processing steps in Figure\,\ref{fig:flowchart}. We focus on a detailed comparison of the maps and power spectra of the kSZ map, following different assumptions regarding the intra-cluster gas.

\subsection{$N$-body simulation and lightcone construction}

We simulate secondary CMB anisotropies using the output from an $N$-body simulation with a box size of $L=1491\,\mpc$, $3200^3$ particles, and a particle mass of $7.43\cdot10^{9}\msol$. This particular run is part of a suite of $\sim 100$ simulations that are being carried out under the Mira-Titan Universe project~\citep{Heitmann:2015xma}. In this paper, we refer to this single instance as the \emph{Mira Universe}, or the \emph{MiraU} simulation. The simulation was performed using the cosmology $N$-body code framework HACC (Hardware/Hybrid Accelerated Cosmology Code; \citealt{Habib:2014uxa}). The initial conditions adopt a WMAP7 cosmology, specifically $\Omega_{\mathrm{DM}}=0.22$, $\Omega_{\mathrm{b}}h^2=0.02258$, $\sigma_8=0.8$, and $h=0.71$. One hundred output snapshots from the simulation are stored between $z=10$ and $z=0$, containing the coordinates and velocities of all particles, requiring $\sim$1\,TB of disk-space per snapshot. In order to identify halos, we run a friends-of-friends (fof) algorithm with a linking length $b=0.168$, where the smallest halo is defined as having 40 particles. 

There are many different ways of defining the mass of a halo. One definition is the \emph{fof mass}, which is simply the mass of all particles that belong to the halo. A different way of defining the halo mass is the \emph{overdensity mass}, which is defined with respect to some reference density. In this work, we define the overdensity mass as
\beq 
M_{200} = \frac{4}{3}\pi R^{3}_{200}\cdot 200\rho_{\mathrm{crit}}(z),
\eeq 
where $R_{200}$ is the radius within which the average density of the halo is 200 times larger than the critical density at redshift $z$. For all halos with more than 1000 particles we compute the overdensity mass $M_{200}$ by measuring the mass overdensity in concentric spheres around the halo center (defined by the gravitational potential minimum).

In this work, we are interested in the kSZ signal from massive clusters at $z<1$, corresponding to the expected redshift range of clusters to be detected with optical surveys such as DES. For this reason, we use only the 29 output snapshots from the simulation between $z=1$ and $z=0$. In order to generate full-sky tSZ and kSZ maps, we first produce two full-sky lightcones up to $z=1$, one from all particles in the simulation, and one from all halos in the simulation for which we have computed $M_{200}$.\footnote{We note at this point that source confusion is not an issue  given the low number density of clusters (5 clusters per square degree). Other sources for potential source confusion could be high-z gas and low-mass clusters. The kSZ power spectrum from high-z structure is however well below current SPT and SPTpol noise levels, so that kSZ analyses with these experiments are not expected to be confusion-limited.}

The lightcones are constructed as follows: First, the original simulation box is replicated twice in all dimensions, completely filling the space up to $z=1$, which corresponds to a comoving distance of $\sim$3300\,Mpc. The location of the observer is chosen to be in the corner $(x,y,z)=(0,0,0)$ of the original box, which is at the same time the center of the total lightcone volume. If a halo from snapshot $s$ crosses the lightcone between the time corresponding to $s$ and the time corresponding to $s+1$, then the time of lightcone-crossing $t_{\mathrm{lc}}$ and the coordinates of the halo on the lightcone $\mathbf{r}_{\mathrm{lc}}$, are computed using the lightcone equation,
\beq
\int_{t_{\mathrm{lc}}}^{t_{\mathrm{final}}} \frac{c\dd t}{a(t)} = |\mathbf{r}_{\mathrm{lc}}|^2,
\eeq
where $t_{\mathrm{final}}$ is the time at the end of the simulation. In order to solve the lightcone equation for each halo in each snapshot, we make the linear approximation $\mathbf{r}_{\mathrm{lc}} = \mathbf{r}(t_s) + \mathbf{v}(t_s) \dd t$. We have checked that the halos in the lightcone follow a distance-redshift relation that is consistent with its theoretical prediction based on the underlying cosmology. The lightcone from particles is constructed in the same way, using the coordinates and velocities of the particles instead of the halos. 

\subsection{Simulated CMB components}
\label{sec:gasmodel}
\begin{figure*}
\centering
\includegraphics[scale=0.36]{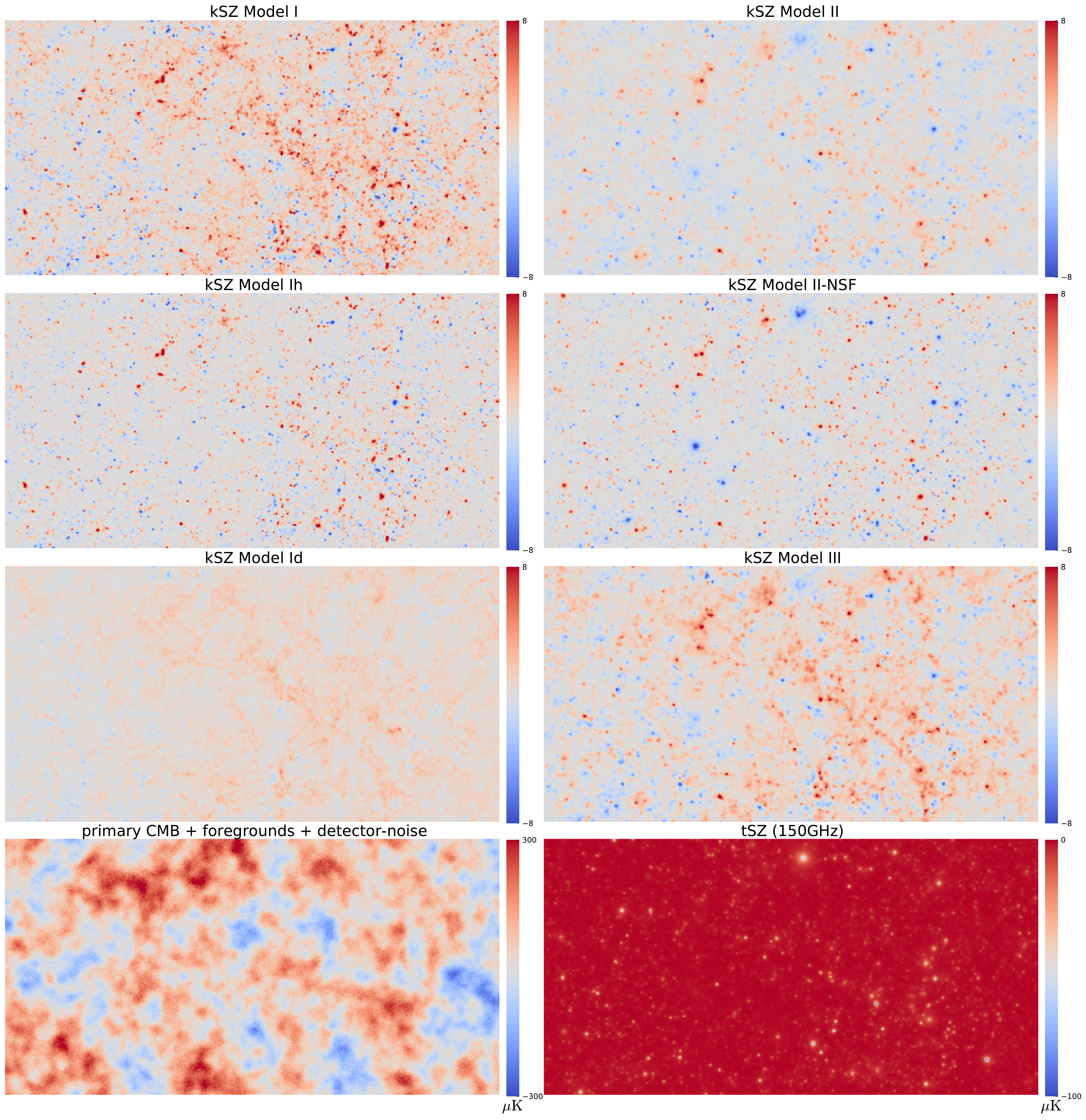}
\caption{\label{fig:szmap} 4x8\,square-degree cutouts from the simulated kSZ maps in different models, the tSZ map, and the noise map, including primary CMB, foregrounds, and detector noise. All maps have been smoothed with a Gaussian beam of FWHM=1arcmin, which corresponds roughly to the SPT instrument beam. By comparing the various kSZ maps we demonstrate the impact of astrophysical effects on the shape of the kSZ profiles. For instance, comparing Model\,II with Model\,II-NSF, which is the same model without star formation and feedback, demonstrates that star formation and feedback make the profile less steep.}
\end{figure*}

\begin{figure*}
\centering
\includegraphics[scale=0.38]{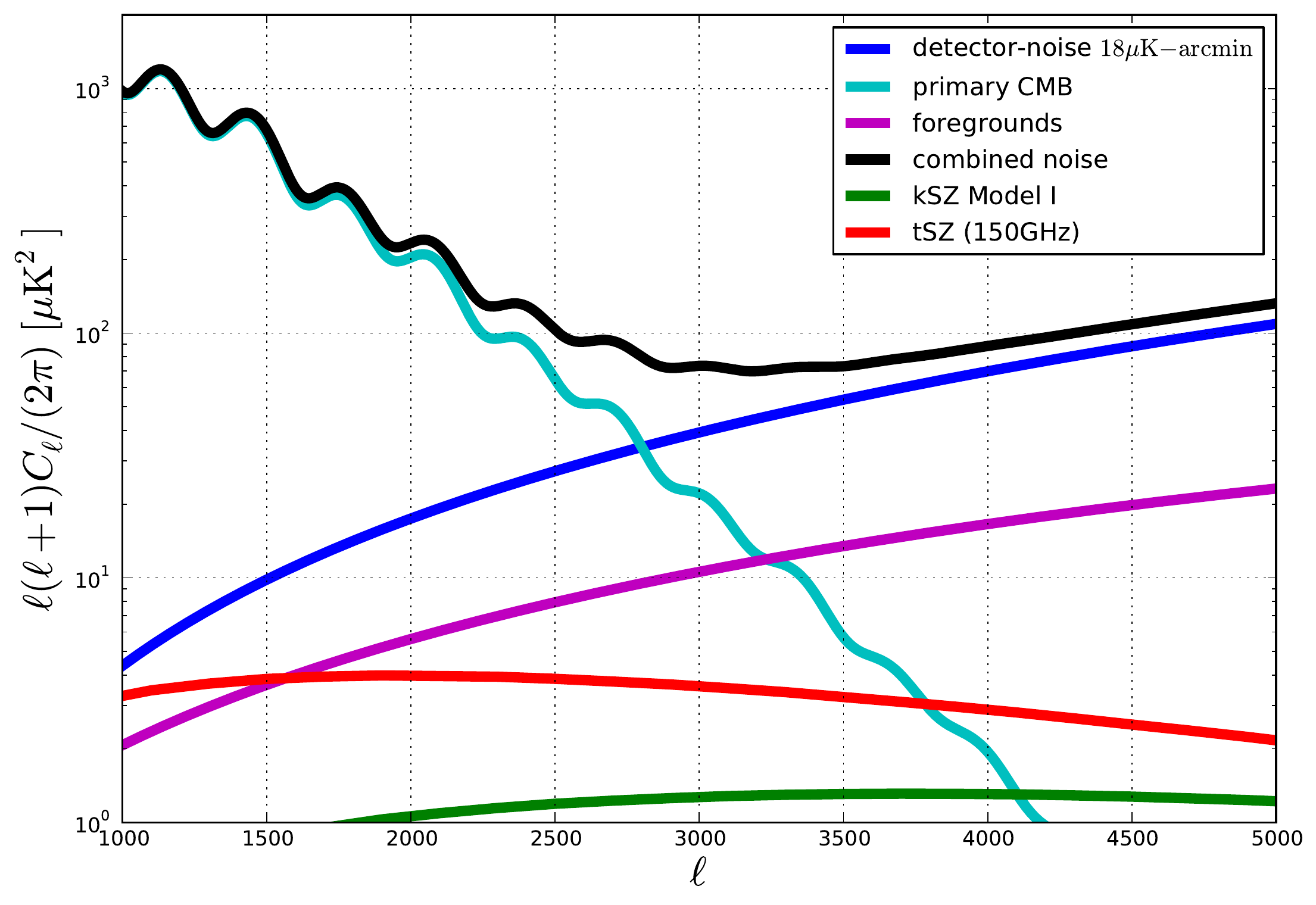}
\includegraphics[scale=0.38]{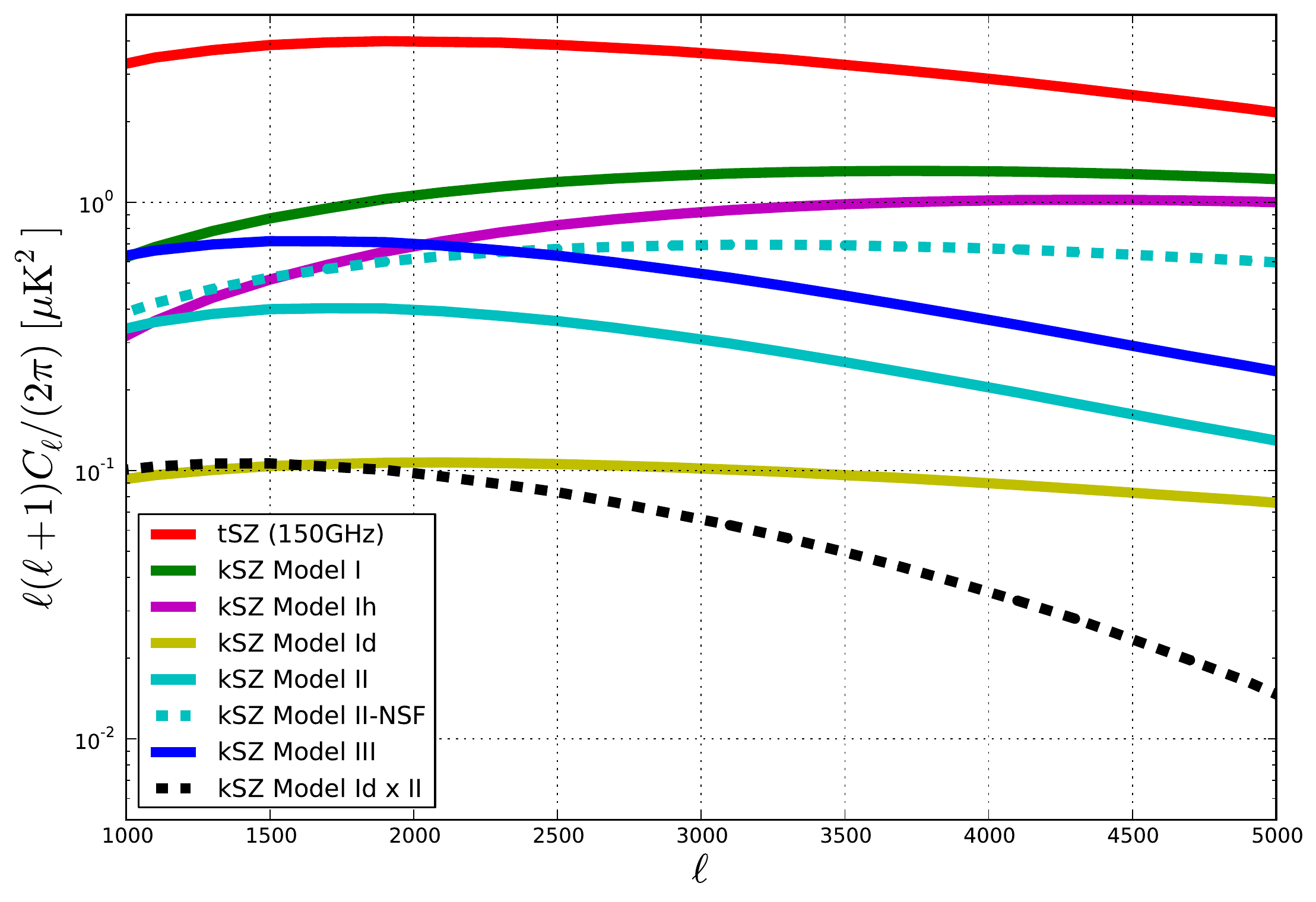}
\caption{\label{fig:spectra} Left: Power spectra of the simulated noise components, including primary CMB, detector noise, and foregrounds. At low-$\ell$, the dominating noise component is the primary CMB, whereas at high-$\ell$ it is the detector noise. Right: Power spectra of simulated SZ components. The tSZ power dominates the kSZ power in all models. The kSZ Model\,I has the most power, as it relies on the assumption that gas traces the dark matter at all scales. Most of the power in Model\,I comes from the halo component (Ih). Star formation and feedback reduce the kSZ power, as can be deduced by comparing Model\,II-NSF to Model\,II. The filament-component and the cluster-component of the kSZ signal are correlated at low-$\ell$, as indicated by the black dashed line. All noise components except the instrument noise have been convolved with a Gaussian beam with FWHM=1arcmin, which corresponds roughly to the SPT instrument beam.}
\end{figure*}

To first order, the kSZ signal along a given line of sight is given by
\beq
\left(\frac{\Delta T}{T}\right)_{\rm{kSZ}} = 
	-\frac{\sigma_{\mathrm{T}}}{c} \int_{\mathrm{los}} \dd l \, n_{\mathrm e} \, v_\mathrm{los}\mbox{ .}
 \eeq
Here, $\sigma_{\mathrm{T}}$ is the Thomson cross-section, $c$ is the speed of light, $\dd l$ is the proper line element, $v_\mathrm{los}$ is the proper line-of-sight velocity, and $n_\mathrm{e}$ is the proper electron number density. We follow the convention that $v_{\mathrm{los}}>0$ corresponds to a cluster moving away from the observer, creating a negative kSZ effect along its line of sight.

\begin{table}
\begin{center}
\begin{tabular}{| p{0.8cm} | p{6.5cm} |}
\hline
kSZ model & brief description \\ \hline \hline
I & kSZ signal derived from all particles in the lightcone, assuming that baryons trace dark matter, using the universal baryon fraction\\ \hline
Ih & halo-component of I\\ \hline
Id & non-halo (diffuse) component of I\\ \hline
II & kSZ signal from clusters, following a gas-prescription based on the Shaw model\\ \hline
II-NSF & II without star formation and feedback\\ \hline
III & II combined with Id\\ \hline
\end{tabular}
\end{center}
\caption{\label{table}}
\begin{tablenotes}
\item Brief summary of the kSZ models studied in this work. See Section  \ref{sec:gasmodel} for details.
\end{tablenotes}
\end{table}

The biggest challenge in modeling the kSZ signal from clusters lies in the uncertainties related to the density profile of the intra-cluster gas. For this reason, we consider several different models built on differing assumptions about the gas distribution; we summarize the essential properties of these models in Table\,\ref{table}. 

The simplest assumption about the gas density is that baryons trace the dark matter at all scales, and this is the assumption we make in what we call {\bf Model\,I}. In this treatment, the baryon density is proportional to the total matter density, where the proportionality factor is simply the cosmic baryon fraction $f_{\mathrm{b}} = \Omega_{\mathrm{b}}/\Omega_{\mathrm{M}}$. In this approximation, we can write the electron number density as 
\beq
n_e = \mu f_{\mathrm b} \rho_m,
\eeq
where $\mu \equiv (1-Y_{\mathrm{He}})/m_{\mathrm{H}}+Y_{\mathrm{He}}/m_{\mathrm{He}}$, with the Hydrogen mass $m_{\mathrm{H}}$, the Helium-mass $m_{\mathrm{He}}$, and the Helium fraction $Y_\mathrm{He} = 0.2477$.
We pixelize the sky using a Healpix{\footnote{http://healpix.sourceforge.net}} pixelization with $\mathrm{Nside}=8192$.
The kSZ signal at a given pixel can then be simply computed from the coordinates and velocities from all particles in the lightcone,
\begin{equation}
\label{eq:eqn_ksz_pix}
\left(\frac{\Delta T}{T}\right)_{\rm{kSZ}} = 
	-\frac{\sigma_{\mathrm{T}}\,f_\mathrm{b}\,\mu\,M_{\mathrm{p}}}{c\,\Omega_{\mathrm{pix}}}\sum_{i}\frac{v_{\mathrm{los},i}}{d_{\mathrm{A},i}^2}\ \mbox{ ,}
\end{equation}
where $M_{\rm p}$ is the particle mass, $\Omega_{\rm pix}$ is the pixel area, $d_{\mathrm{A},i}$ is the angular diameter distance to particle $i$, and the sum is taken over all particles in that pixel. This method introduces considerable shot noise at low redshifts. For this reason, we restrict the resulting map to the redshift range $0.1<z<1$. Our kSZ Model\,I corresponds to the model chosen in \cite{Li:2014mja}, however with a higher angular resolution, larger redshift range, and more output snapshots used for interpolating the lightcone.

The assumption made in Model\,I, i.e., that baryons trace the dark matter, is somewhat reasonable on large scales. On cluster scales, however, the gas profile is different from the dark matter profile due to a number of physical effects. In kSZ {\bf Model\,II}, we implement a gas prescription for the intra-cluster gas, following the model introduced by \citeauthor{Shaw:2010mn} (\citeyear{Shaw:2010mn}; hereafter the Shaw Model). Below we briefly summarize this model, with a focus on the free parameters therein. We refer the reader to \cite{Shaw:2010mn} for the computational details.

The Shaw model, like the earlier models by \cite{Bode:2009gv} and \cite{Ostriker:2005ff}, assumes that the gas inside the cluster initially follows the dark matter density distribution, which is modeled as a Navarro-Frenk-White (NFW) profile~\citep{Navarro:1996gj}, and then rapidly rearranges itself into a state of hydrostatic equilibrium inside the halo, following a polytropic equation of state with polytropic index $\Gamma=1.2$. The model makes further corrections to the gas profile by taking into account star formation, feedback, and non-thermal pressure. Star formation is modeled using the empirical relation from \cite{Giodini:2009qf}
\footnote{We note that the star-formation model by \cite{Giodini:2009qf} predicts a larger amount of star-formation as compared to other models proposed in the literature (see, e.g.\ \citealt{Leauthaud:2011gw, Lin:2004hw, Budzynski:2013obt, Zu:2015cpa}). Throughout this work we will assume the model by \citeauthor{Giodini:2009qf}, since this is the model that was used in the analysis by \citet{Shaw:2010mn}, which produces a tSZ power spectrum that is consistent with SPT constraints. Exploring the uncertainty in our results due to the specific choice of the star-formation model is beyond the scope of this work.}. 
The free parameters in the Shaw model are as follows:
\bim
\item The amount of energy transfer from the dark matter to the gas during major mergers is modeled with one parameter, $\epsilon_{\mathrm{DM}}$.
\item The amount of energy feedback from supernovae and active galactic nuclei is modeled with one parameter, $\epsilon_{\mathrm f}$.
\item The amount and redshift dependence of non-thermal pressure support due to random gas motions are modeled with the two parameters, $\alpha_0$ and $\beta$.
\eim
Here, we adopt the fiducial model from \cite{Shaw:2010mn}, i.e.,\ $\epsilon_{\mathrm f} = 10^{-6}$, $\epsilon_{\mathrm{DM}}=0.05$, $\alpha_0=0.18$, and $\beta=0.5$, which was shown to produce a tSZ power spectrum that is in agreement with observational constraints from SPT \citep{lueker10}. We produce a full-sky Model\,II-kSZ map as follows. For each halo in the lightcone, we assign a concentration based on its mass $M_{200}$, following the concentration-mass relation from \cite{2013ApJ...766...32B}, which has been recently shown to be in good agreement with lensing measurements of cluster profiles (\citealt{2013ApJ...769L..35O}, \citealt{2015ApJ...806....4M}). Then, we assign to each Healpix pixel within $3\theta_{200}$ of the halo center the kSZ amplitude at that pixel, given the distance of the pixel to the halo center. Additionally, in order to test how much the kSZ signal depends on star formation and feedback, we create a variant model, Model\,II-NSF, which is the same as Model\,II, but without star formation and feedback. 

Model\,II only takes into account the cluster-component of the kSZ, and neglects completely the contributions from filaments and the intergalactic medium. We therefore construct another model, which we call {\bf Model\,III}, by adding these missing contributions into Model\,II. We do this by dividing all particles in the lightcone into two groups, depending on whether or not they were identified as halo particles by the fof algorithm. This results in two separate kSZ models, which we call Model\,Ih and Model\,Id, respectively. Model\,III is then simply the sum of the kSZ maps produced from Model\,II and Model\,Id
\footnote{Note that in Model\,III there is some overlap between the region within which we assign the Shaw profile (i.e., within $3\theta_{200}$ of the cluster center) and the filament component from Model\,Id. This overlap is however not a problem because the Shaw profile falls off steeply (most of the signal is contained within $\theta_{200}$).}.

Note that, by construction, Model\,III does not include the kSZ component from halos with 40-1000 particles, i.e.\ halos with fof-masses of $3\cdot10^{11}-7.4\cdot10^{12}\msol$. This is however not a problem for the purpose of this work, since here we are interested in the kSZ signal from groups and clusters with masses above $10^{13}\msol$. Furthermore, this missing component amounts to only $4.2\%$ of the total lightcone mass, whereas halos with >1000 particles account for $9.6\%$, and $86.2\%$ comes from non-halo components. Nevertheless, we can easily take that missing component into account by combining the map produced in Model\,II with the Model\,Id map, the latter scaled by a factor of $(86.2+4.2)/86.2 = 1.049$. We refer to this map as Model\,IIIs. Models\,III and IIIs are thus almost the same, and we find essentially identical results (both for the power spectrum and the pairwise kSZ signal) for these two models. We therefore quote in the remainder of this work only the results for Model\,III. 

We stress at this point that none of our models takes into account the contribution to the kSZ power from low-mass halos ($<10^{13}\msol$), which can be of significant amplitude (see e.g.\ \citealt{Singh:2015cua}). In our Model\,IIIs we make the assumption that these halos have been `smeared out' along the filaments. As stated above, this is not an issue for the analysis of the cluster component of the kSZ signal, which is our primary concern in this work. Our maps are however not sufficient for studies of the kSZ signal from low-mass halos.

The tSZ signal at frequency $\nu$ is at first order given by $\Delta T/T_{\mathrm{CMB}} = f(x_{\nu})y$, where $f(x_{\nu})=x_{\nu}(\coth(x_{\nu}/2)-4)$, $x_{\nu}=h\nu/(k_{\mathrm{B}}T_{\mathrm{CMB}})$, and $y$ is the dimensionless Compton-$y$ parameter, which can be expressed as the line-of-sight integral of the electron-pressure $P_e$ \citep{Sunyaev:1980nv},
\beq
y = \frac{\sigma_{\mathrm T}}{m_e c^2}\int \dd l P_e(l).
\eeq
We produce a full-sky tSZ map by painting a tSZ profile, following the Shaw model, into a blank Healpix map at the location of each halo in the lightcone. 

To help demonstrate the differences between the kSZ models we present cutouts of maps generated from each of the models in Figure\,\ref{fig:szmap}, as well as power spectra of the full maps in the right-hand panel of Figure\,\ref{fig:spectra}. Here we note a few important features that can be derived from those figures:

\begin{itemize}

\item The difference between Model\,I and Model\,Ih (top left panels in Figure\,\ref{fig:szmap}) is due to the removal of the diffuse kSZ component in the latter, which corresponds to a removal of large-scale power. While excision of the diffuse component removes $\sim$90\% of the mass, this only reduces the power at $\ell=3000$ (right hand panel Figure \ref{fig:spectra}) by $\sim 30\%$.

\item The difference between Model\,Ih and Model\,II is due to the different modeling of the intra-cluster gas. In Model\,II, star formation and feedback make the individual cluster profiles less sharp and thus remove power at small scales.  This also becomes clear when comparing Model\,II to Model\,II-NSF, which is Model\,II without star formation and feedback. Model\,II-NSF looks much more like Model\,Ih, i.e., the cluster profiles are more NFW-like. Model\,II has consequently less power on small scales as compared to Model\,II-NSF (e.g., at $\ell=3000$ there is 50\% less power in Model\,II).

\item The reason for the difference between Model Ih and Model II-NSF is the lack of substructure in the latter. Model II-NSF assumes that the gas profile follows a spherical NFW profile and then re-arranges into hydrostatic equilibrium. In reality however, halos have substructure, including velocity substructure, which is included in Model Ih.

\item The difference between Model\,II and Model\,III is the inclusion of the diffuse component in Model\,III, which adds power at large scales. 

\item Finally we note that at large scales, the cluster-kSZ component and diffuse kSZ component are correlated, as can be seen from the black dashed line in Figure\,\ref{fig:spectra}. This is expected, as groups and clusters form in filaments and thus trace their large-scale distribution.

\end{itemize}

In addition to the SZ maps, we also simulate the primary CMB anisotropies, foregrounds, and CMB detector noise; all of these components are considered to be ``noise'' for our analysis of the kSZ signal. For generating the angular power spectrum of the primary CMB we run CAMB\footnote{http://camb.info} \citep{lewis00}, using the same cosmological parameters as those used for the {\em MiraU} simulation. We model three types of foregrounds: Poisson noise from both dusty star-forming galaxies and radio galaxies, as well as the clustered component from the cosmic infrared background (CIB). All of these are modeled as random Gaussian realizations using the best-fit model from recent SPT measurements presented in \cite{George:2014oba}. We model the detector noise as white noise with a noise level of 18\,\muk-arcmin, corresponding to the SPT-SZ 150\,GHz-channel. We show the power spectra of the noise components in the left panel of Figure\,\ref{fig:spectra}. At low multipoles the dominant noise component is the primary CMB, whereas at high multipoles the noise is dominated by instrument noise. The kSZ signal is far sub-dominant to other contributors to the millimeter-wave sky and is thus challenging to isolate.


\section{The pairwise kSZ signal}
Galaxy clusters move on average towards each other due to their mutual gravitational attraction and, as explained above, this process creates a distinct pattern in the CMB that we call the pairwise kSZ signal. Extracting this signal from CMB data requires two post-processing steps: First, filtering of the CMB map in order to optimally reconstruct the (sub-dominant) kSZ signal, and second, applying an estimator that uses the locations of clusters in order to measure the pairwise kSZ signal created by those clusters. In this section, we describe these two post-processing steps in detail, and present the resulting pairwise kSZ signal for our various kSZ models. We explore several observational limitations that may bias the recovered signal and combine these findings to present realistic forecasts for the kSZ signal to be obtained from current and next generation optical and CMB surveys. 

\subsection{Filtering methods}
Here we adopt two filtering choices for extracting the kSZ signal from the simulated maps: the matched filter and the compensated top-hat filter. We compare the results derived from these filtering choices later in this section. 

\subsubsection{The matched filter}
\label{sec:mf}
Given a signal with known profile $\tau$ embedded into a noise background $N$ with known spectral shape, a filter can be constructed that has minimum variance, called the matched filter. In the context of the kSZ signal, the matched filter was first discussed in \cite{Haehnelt:1995dg}. In harmonic space, the matched filter can be written as (\citealt{McEwen:2006ke})
\beq
\Psi_{\ell m} = \frac{\tau^{\star}_{\ell m}/N_{\ell}}{\sum_{\ell m} |\tau^2_{\ell m}|/N_{\ell}}.
\eeq
If the filter profile $\tau$ is azimuthally symmetric, i.e., $\tau_{\ell m}=0~\forall~m>0$, then the matched filter can be further simplified to
\beq
\Psi_{\ell m} = \sqrt{\frac{4\pi}{\ell(\ell+1)}}\frac{\tau^{\star}_{\ell 0}/N_{\ell}}{\sum_{\ell} |\tau^2_{\ell 0}|/N_{\ell}},
\eeq
and this is the form used here.

Under the assumption that baryons trace the dark matter, the natural choice for the profile $\tau$ is the projected NFW profile, convolved with the beam of the CMB experiment, $\tau=\tau_{\mathrm{pNFW}}*B$, where we approximate the beam $B$ as a Gaussian function with full width at half maximum (FWHM)~$=1$', and the projected NFW profile is given by \citep{1996A&A...313..697B}:
\beq
\tau_{\mathrm{pNFW}}(x)=\frac{A}{x^2-1}F(x),
\eeq
with
\begin{equation}
\label{eq:profile}
	F(x) = 
	\begin{cases}
	1-{2 \over \sqrt{1-x^2}}\ \mbox{tanh}^{-1} \sqrt{{1-x\over x+1}}  &  0<x<1 \\
	0 & x=1 \\
	1-{2 \over \sqrt{x^2-1}}\ \mbox{tan}^{-1} \sqrt{{x-1\over x+1}} & x>1, \\
	\end{cases}
\end{equation}
where $A$ is the normalization of the profile, and $x=\theta/\theta_{\rm S}$, where $\theta_{\rm S}$ is the angle subtended by the scale radius $R_{\rm S}$ of the NFW profile. Alternatively, the profile can also be parameterized with $x=c_{200}\theta/\theta_{200}$, where $c_{200}$ is the concentration of the cluster measured at $R_{200}$, and $\theta_{200}$ is the angle subtended by $R_{200}$. We normalize the smoothed projected NFW profile to unity at $x=0$. We obtain the harmonic coefficients of the smoothed projected NFW profile by first painting that profile onto the North Pole of a blank Healpix map of Nside=8192, and then performing a harmonic transform of that map.

\subsubsection{The compensated top-hat filter}
The compensated top-hat filter can be expressed as
\beq
\tau(\theta)=\begin{cases}
	1 & 0<\theta<\theta_{\rm F} \\
	-1 & \theta_{\rm F}<\theta<\theta_{\rm out} \\
	0 & \mathrm{else},
\end{cases}
\eeq
where $\theta_{\rm F}$ is the characteristic filter scale and $\theta_{\rm out}=\sqrt{2}\theta_{\rm F}$. The compensated top-hat filter subtracts from the average temperature inside a disc the average temperature in the surrounding ring with equal area. Thus, the compensated top-hat filter effectively removes noise on all scales that are larger than the filter scale.

\subsection{The pairwise estimator}
The pairwise estimator was originally derived in \cite{Ferreira:1998id} as the optimal estimator for reconstructing the mean pairwise velocity of objects, given only their line-of-sight peculiar velocities. Peculiar velocities are, however, difficult to measure. In our case, the line-of-sight velocity is replaced by its proxy, the kSZ temperature signal of a given cluster. In this case, the pairwise estimator can be written in terms of CMB temperatures instead of line-of-sight velocities (e.g., \citealt{Hand:2012ui}),
\beq
\hat{T}_{\mathrm{kSZ}} = \frac{ \sum_{ij} c_{ij}\,T_{ij}}{\sum_{ij} c^2_{ij}},
\label{p_est}
\eeq
where $T_{ij}$ is the difference in filtered temperature values at the cluster locations, and $c_{ij}$ are geometric weights given by 
\beq
c_{ij} \equiv {{\bf\hat r}_{ij}}\cdot\frac{{\bf\hat r}_i + {\bf\hat r}_j}{2} = \frac{(r_i - r_j)(1+\cos\theta)}{2\sqrt{r_i^2 + r_j^2 - 2r_ir_j\cos\theta}}.
\eeq 
Here, ${\bf\hat r}_i$ is the unit vector pointing to cluster $i$, ${\bf\hat r}_{ij}$ is the unit vector pointing from cluster $i$ to cluster $j$, $r_i$ is the comoving distance of cluster $i$, and $\theta$ is the angular separation between the two clusters. The weights $c_{ij}$ are designed to assign high weight to cluster pairs oriented along the line of sight and low weight to cluster pairs oriented perpendicular to the line of sight, as we expect a larger kSZ contribution from the former than from the latter.

We compute the filtered temperature difference as $T_{ij} = (T_i - \mathcal{T}(z_i)) - (T_j - \mathcal{T}(z_j))$, where $\mathcal{T}(z_i)$ is defined as
\beq
\mathcal{T}(z_i) = \frac{ \sum_k T_k\,w(z_i,z_k)  }{ \sum_k w(z_i,z_k) },
\eeq
where sums are taken over the whole cluster sample, and $w(z_i,z_k)=\exp({-(z_i-z_k)^2/(2\sigma^2_{\mathrm K}}))$. The purpose of the function $\mathcal{T}(z)$ is to remove redshift-dependent noise from the sample, which could be confused with the pairwise kSZ signal. For instance, the tSZ signal introduces redshift-dependent noise due to the evolution in average cluster mass. In agreement with \cite{Hand:2012ui} and \cite{Ade:2015lza} we choose a kernel size of $\sigma_{\mathrm K}=0.01$.

In the comparisons below, we create 10 distance-bins with equal width from 0 to 200\,Mpc and compute the pairwise kSZ signal in each bin. The error bars are obtained by bootstrapping on the cluster sample. From the bootstrap samples we compute the covariance matrix $\mathrm{cov}(b,b^{\prime})$ between the separation bins $b$. Following \cite{Keisler:2012eg}, we estimate the signal-to-noise ratio (SNR) as $\mathrm{SNR}=\sqrt{\chi^2}$, which is a reasonable approximation for high significance levels and small numbers of degrees of freedom.

\subsection{Comparison of the kSZ models}
\begin{figure*}
\centering
\includegraphics[scale=0.60]{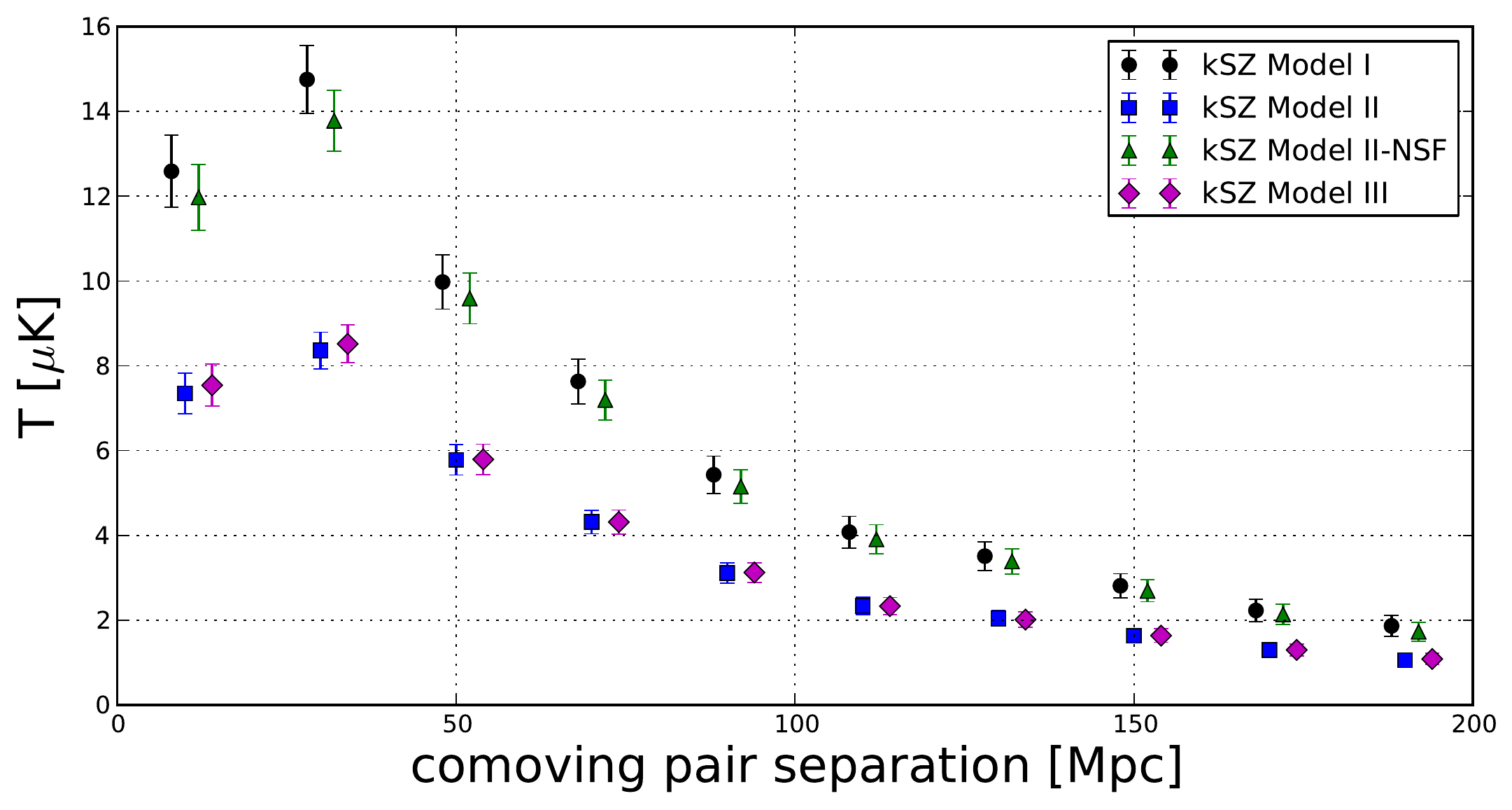}
\caption{\label{fig:comp_models} Pairwise kSZ signal using different kSZ models, without noise components. kSZ Model\,I relies on the assumption that gas traces the dark matter, without taking into account astrophysical effects, while kSZ Model\,II follows a gas prescription for the intra-cluster gas. The significant difference between Model\,I and Model\,II is caused by the change in the gas fraction and the flattening of the gas profile caused by star formation and feedback. This becomes clear when considering that Model\,II-NSF, which is Model\,II without star formation and feedback, gives roughly the same amplitude as Model\,I. Finally, Model\,III and Model\,II show roughly the same signal. This can be explained by the fact that the filament-component of the kSZ, which is included in Model\,III, is not correlated with the cluster-component of the kSZ at small scales.}
\end{figure*}

For our model comparisons we consider one fixed cluster catalog, which we call our baseline-model, with the mass-cuts $10^{14}\msol<M_{200}<3\cdot10^{14}\msol$ and the redshift-cuts $0.2<z<1.0$. In optical surveys one usually works with a mass-proxy called the cluster \emph{richness}, which corresponds roughly to the number of member galaxies in the cluster; the lower mass-cut we consider here, $M_{200}=10^{14}\msol$, corresponds to a richness of $\lambda\sim20$ (\citealt{Rykoff:2011xi}). The upper mass-cut is chosen to remove the largest tSZ sources and maximize the signal-to-noise. We assume an overlap with a CMB map of 2500 square-degrees. There are $\sim 11,700$ clusters in our baseline-model, which corresponds to $\sim5$ clusters per square-degree. The CMB detector noise is modeled as white noise with a noise level of $18\,\mu$K-arcmin. Our baseline-model thus represents roughly the combination of the 5-year DES data and the SPT-SZ CMB data in the 150\,GHz channel.

As explained above, the projected NFW profile can be parametrized with two free parameters, $c_{200}$ and $\theta_{200}$. We assign to each cluster a concentration based on its mass, following the concentration-mass relation found by \cite{2013ApJ...766...32B}, and use the mean values of our baseline cluster sample, $\bar{c}_{200}=3.5$ and $\bar{\theta}_{200}=2.9^{\prime}$, as filter parameters for the matched filter. In principle, we could also adjust the filter scale to each cluster individually, which is computationally much more expensive. Another strategy is to divide the cluster sample into bins of $\theta_{200}$ and design a matched filter for each bin. We have checked that the results with this method are not significantly different from the ones obtained with one matched filter, using the average $\theta_{200}$ from the whole sample.

In Figure\,\ref{fig:comp_models} we show the pairwise kSZ signal as a function of comoving separation for different kSZ models. For clarity, we show here the results from the kSZ-only maps without any noise added, i.e.,\ the scatter corresponds to the intrinsic scatter in the kSZ only. Qualitatively, the pairwise signal shows the same behavior in the different models: The signal decreases with increasing pair separation, as cluster pairs at small separations feel their mutual gravitational pull, and cluster pairs at large separations move with the Hubble flow. At very small separations (of a few Mpc) the pairwise velocity becomes non-linear. In that non-linear regime, clusters are not always moving towards each other due to non-linear effects, i.e. virialization, collisions, and mergers. These effects lead to a suppression of the pairwise velocity in the first separation bin. Thus, the signal peaks at the second bin.

Quantitatively, the results for Model\,I and II are significantly different. In particular, we find that the amplitude of the pairwise kSZ signal is larger by a factor of $\sim$2 in Model\,I as compared to Model\,II. This difference can be explained by the different treatment of the intra-cluster gas in the two models. Model\,I neglects star formation and feedback, which reduce the amount of gas inside the cluster and make the gas profile shallower, i.e.,\ less NFW-like. These effects decrease the kSZ amplitude, which is proportional to the projected gas density. This becomes further apparent when considering that Model\,II-NSF, which is  Model\,II without star formation and feedback, produces roughly the same amplitude as Model\,I. Finally, Model\,III is consistent with Model\,II. This suggests that the filaments do not add any signal, which can be explained by the fact that the cluster-kSZ component and the filament-kSZ component are correlated only at large scales, but not at small scales (see Figure\,\ref{fig:spectra}).

Now including the primary CMB, foregrounds, and instrumental noise, we estimate the signal-to-noise ratio in this data scenario as $\mathrm{SNR}=8-13$, depending on the model, where the lower bound corresponds to kSZ Model\,II/III, and the upper bound corresponds to Model\,I/II-NSF. Our results stress that the SNR-forecast depends strongly on the treatment of the intra-cluster gas. For comparison, \citet{Keisler:2012eg} estimate a significance level of 12.5$\sigma$ for a similar data scenario.

We note that our results, including the SNR forecasts, are derived under the assumption of a WMAP7 cosmology. Different results are expected to be found for different cosmologies, in particular because of the strong dependence of the clusters counts on $\sigma_8$. A detailed investigation of the parameter dependence is beyond the scope of this work.

\subsection{Dependence on the mass threshold}

\begin{figure}
\centering
\includegraphics[scale=0.40]{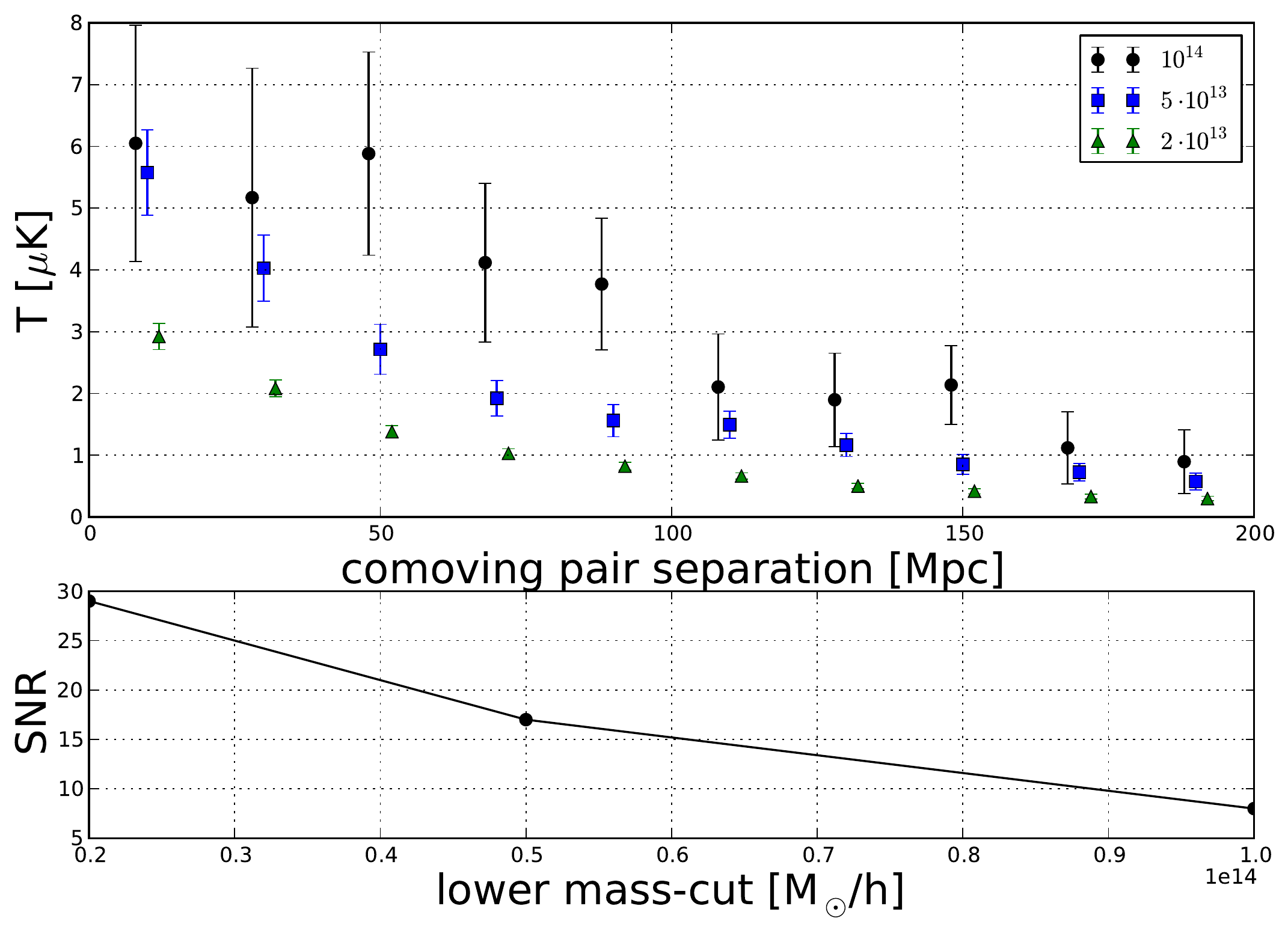}
\caption{\label{fig:var_masscut} Pairwise kSZ temperature signal as a function of comoving cluster separation, for varying lower mass-cuts. The upper mass cut is fixed at $3\cdot10^{14}\msol$. The lower panel shows the SNR as a function of the lower mass-cut. Decreasing the mass cut results in a lower amplitude of the signal because lower-mass clusters produce a lower kSZ signal. At the same time, the SNR increases because of the larger sample size. Note that here we are not taking into account systematic uncertainties such as redshift errors and mis-centering, which might increase for clusters with lower richness.}
\end{figure}

We next investigate the impact of the choice of the lower mass cut on the performance of the pairwise estimator. In our baseline model we choose the lower mass cut $M_{200}=10^{14}\msol$, which corresponds to a richness of $\lambda \sim 20$, the expected high-purity selection threshold for DES clusters. Lower mass thresholds might, however, be achievable in exchange for lower purity of the cluster sample. We find that lowering the threshold from $10^{14}\msol$ to $5\cdot10^{13}\msol$ roughly quadruples the sample size, from 11,700 to 50,000 objects, and the SNR roughly doubles from $\sim$8 to $\sim$17, using kSZ Model\,III. Further decreasing the threshold to $2\cdot10^{13}\msol$ results in a sample of 240,000, and an SNR of $\sim$29. The overall amplitude of the signal decreases with decreasing mass threshold (see Figure\,\ref{fig:var_masscut}), which is expected as lower-mass clusters produce a smaller kSZ signal (see also \citealt{Dolag:2013hj}).

We stress that the relation between the SNR and the mass threshold shown here only holds for a cluster sample where the redshifts and centers are perfectly known. In practice however, photometric redshift errors and mis-centering might increase with decreasing cluster richness, and therefore a different relation between the SNR and the mass threshold might be expected. We discuss the effects of redshift errors and mis-centering in detail in the following.

\subsection{Systematic errors}
\subsubsection{Bias from photometric redshift errors}

\begin{figure}
\centering
\includegraphics[scale=0.40]{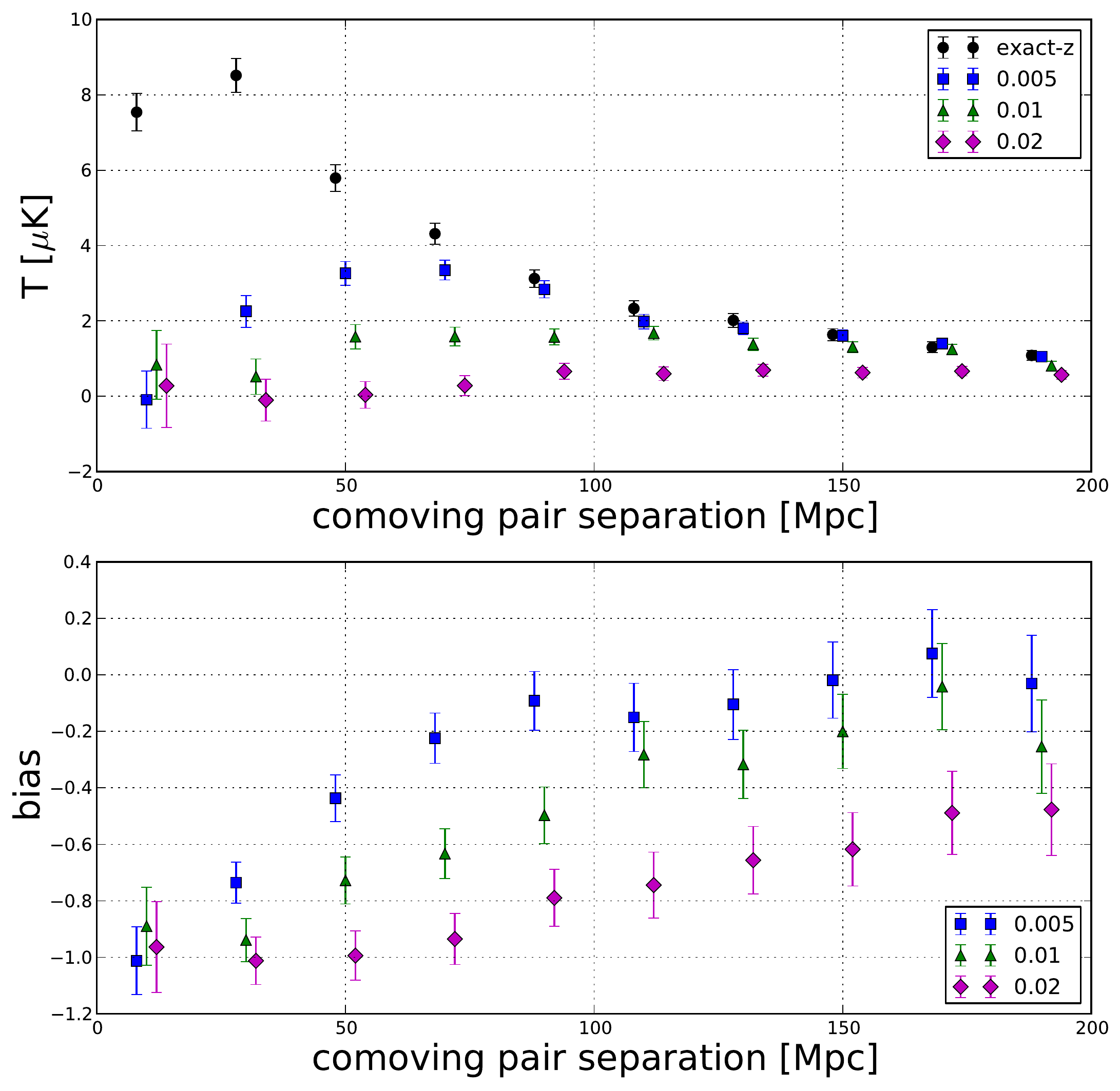}
\caption{\label{fig:vary_sigmaz} Top: Pairwise kSZ signal as a function of comoving pair separation for different values of $\sigma_{z,0}$, without other noise components. Bottom: Bias from redshift errors, as a function of comoving separation. Increasing $\sigma_{z,0}$ leads to a stronger bias. In all cases, we find an almost complete suppression of the signal at low separations.}
\end{figure}

For a cluster pair with small separation, redshift errors can confuse the information about which cluster partner is closer and which is farther away from the observer. A redshift error of $\Delta z$ translates into an error in the comoving distance $\Delta r = c\Delta z/H(z)$; for instance, a redshift error of 0.01 translates into an error in the comoving distance of around 25\,Mpc at $z=1$, and 38\,Mpc at $z=0.2$, and thus dilutes the pairwise kSZ signal below those scales. This leads to a limitation of the detection significance, as was already shown in the analysis in \cite{Keisler:2012eg}.

Here, we wish to quantify the bias introduced by redshift errors. We model redshift errors by adding to each cluster redshift $z$ an error $\Delta z$, which is drawn from a Gaussian probability distribution with zero mean and standard deviation $\sigma_z=\sigma_{z,0}(1+z)$, where $\sigma_{z,0}$ is the free parameter in our model. For a given $\sigma_{z,0}$ we compute the bias $B(d,\sigma_{z,0})=(T(\sigma_{z,0})-T_{\mathrm{exact}})/T_{\mathrm{exact}}$ as a function of cluster separation, where $T_{\mathrm{exact}}$ is the pairwise kSZ estimate using exact redshifts. We neglect here any richness-dependence of the photometric redshift errors, which might be present in observational data. For our purpose, which is a demonstration of the relation between $\sigma_{z,0}$ and the bias in the kSZ amplitude, the simple model considered here is sufficient. 

We vary $\sigma_{z,0}$ from 0.005 to 0.02 (see Figure\,\ref{fig:vary_sigmaz}), which corresponds roughly to the range of redshift errors that is expected in a photometric redshift survey. For this study, we ignore the error contributions from the primary CMB, foregrounds, and the tSZ signal, i.e.,\ the error bars refer solely to the variance in the kSZ and the redshift errors. As expected, we find that $\sigma_{z,0}$ introduces a negative bias to the pairwise kSZ estimate, which becomes as large as $-1$ at the lowest separation bins. In other words, redshift errors can completely remove the signal at the smallest separation bins. At a separation of 100\,Mpc, the bias can be between $-0.1$ and $-0.8$, depending on the value of $\sigma_{z,0}$. These results are obtained using kSZ Model\,III but are model independent.

Redshift errors not only bias the kSZ amplitude, they also reduce the detection significance of the pairwise kSZ signal. With $\sigma_{z,0}=0.01$, which corresponds roughly to the expected redshift errors of clusters detected in DES \citep{Abbott:2005bi}, we estimate a signal-to-noise ratio of $\sim$4-7 in our baseline-model, depending on the kSZ model, where the lower bound comes from kSZ Model\,III and the upper bound from Model\,I. 

We have thus demonstrated that redshift errors both introduce a significant bias into the amplitude of the measured pairwise kSZ signal, and limit the detection significance. Nevertheless, as we have shown here, a strong detection of the pairwise kSZ signal is expected even in the presence of redshift errors with data from SPT and DES, and we thus confirm the findings of \cite{Keisler:2012eg}.

\subsubsection{Mis-centering bias}

\begin{figure}
\centering
\includegraphics[scale=0.4]{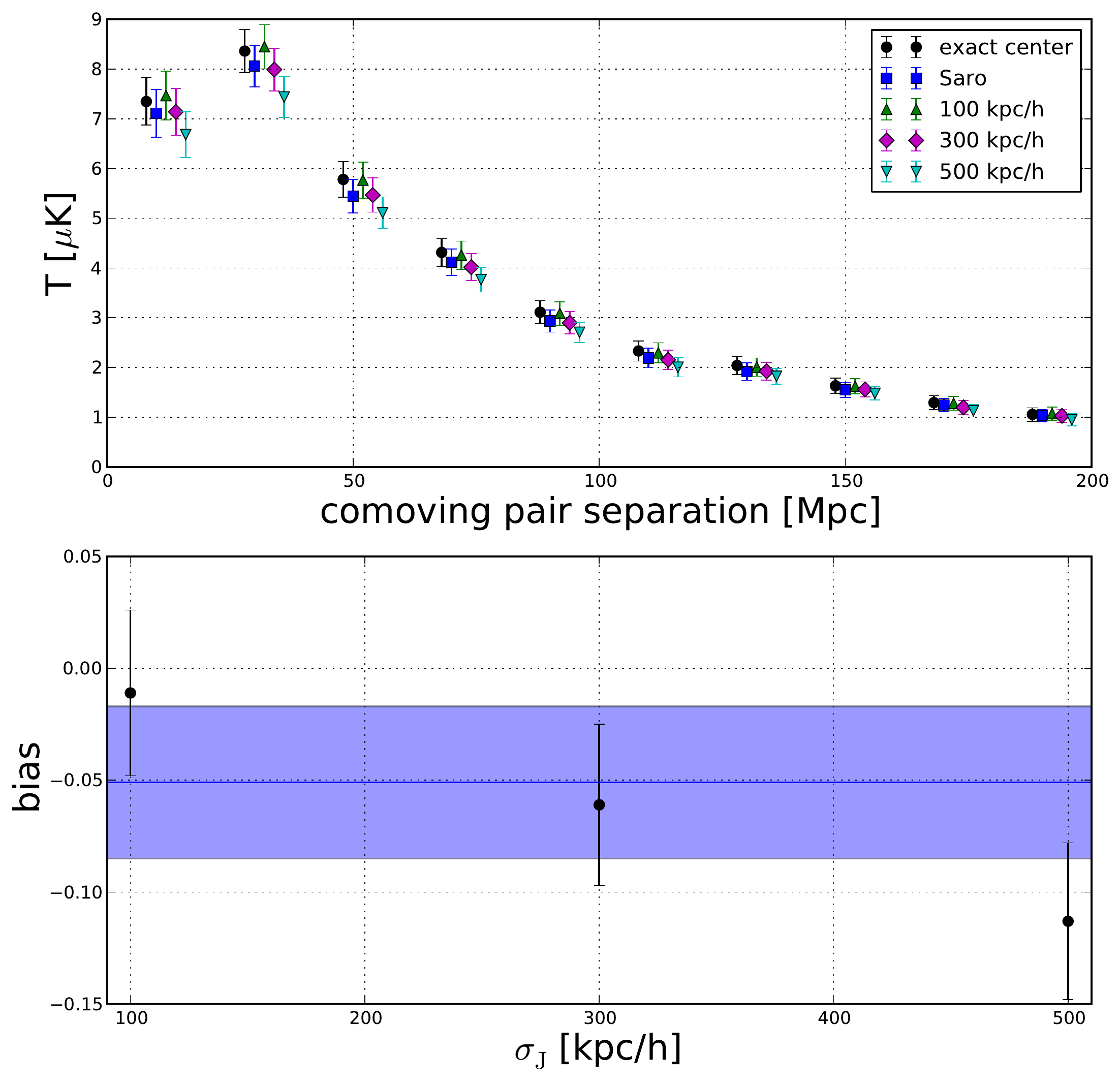}
\caption{\label{fig:misc} Top: Pairwise kSZ signal as a function of comoving separation for different mis-centering models, without noise components. We consider two mis-centering models, the Saro model and the Johnston model with parameters $f_{\mathrm{J}}=0.75$ and varying values for $\sigma_{\mathrm J}$ of  100, 300, and 500\,kpc/h. Bottom: mean mis-centering bias $\bar{B}$ as a function of $\sigma_{\mathrm J}$ in the Johnston model. For comparison, the mis-centering bias in the Saro model is indicated by the blue band. Mis-centering biases the amplitude over all separation bins. This effect is as large as -11\% for the models considered here.}
\end{figure}

In observations, we do not know the ``true'' center of a cluster---which in our simulations we have defined as the location of the gravitational potential minimum. A commonly used proxy for the cluster center is the location of the brightest galaxy in the cluster (BCG). However, due to mis-identification of cluster members it can happen that the galaxy labeled as BCG is not actually the brightest galaxy in the cluster. Furthermore, even if the BCG is labeled correctly, it is not always at the location of the potential minimum (or, more relevant for studies of the kSZ signal, the projected center of the electron distribution; here we assume these coincide). For these reasons, there can be an offset $R_{\mathrm{off}}$ between the cluster center estimated from the BCG and the true cluster center.

Here, we consider two different models for the mis-centering distribution. Our first model follows \citet{Johnston:2007uc}, which we hereafter call the Johnston model. In this model we assume that the cluster sample contains a fraction $f_{\mathrm J}$ of clusters with exact centers, corresponding to the case where the BCG is actually close to the potential minimum, and a fraction $1-f_{\mathrm J}$ of clusters with an off-centering that follows a 2D-Gaussian distribution with width $\sigma_{\mathrm{J}}$. The Johnston model has thus two free parameters, $f_{\mathrm J}$ and $\sigma_{\mathrm J}$. \citet{Lin:2004hw} compare the location of the BCG to the X-ray-determined center for 93 groups and clusters, and find that the BCG is located very close to the X-ray-determined center (within $0.06 R_{200}$) in 75\% of the cases. We thus choose $f_{\mathrm J}=0.75$, and vary $\sigma_{\mathrm{J}}$ from $0.1$ to $0.5\mpc$. Mis-centering by more than $0.3\mpc$ is however very rare \citep{0004-637X-783-2-80}.

Our second mis-centering model is adopted from \citet{Saro:2015lqu}, and we refer to this model as the Saro model. \citet{Saro:2015lqu} compare the positional offset of the BCG in clusters detected in DES with the center derived from the tSZ signal, using data from SPT. They find that the mis-centering distribution can be modeled as a 2-component Gaussian distribution,
\beq 
p(x) = 2\pi x \left( \frac{\rho_0}{2 \pi \sigma_0^2}e^{{-x^2/2 \sigma_0^2}} + \frac{1- \rho_0}{2 \pi \sigma_1^2}e^{{-x^2/2 \sigma_1^2}} \right) 
\eeq
where the mis-centering width is given in terms of $x=r/R_{500}$. This model has thus 3 free parameters, $\rho_0$, $\sigma_0$, and $\sigma_1$. The authors show that the sample is well characterized as a large population ($\rho_0=0.63$) of clusters with small mis-centering ($\sigma_0=0.07$), and a subdominant population with larger mis-centering ($\sigma_1 = 0.25$). Note that the Johnston model corresponds to the Saro model in the limit $\sigma_0\rightarrow0$.

For both mis-centering models we measure the bias at comoving separation bin $b$, $B(b)=(T_{\mathrm{misc}}(b)-T_{\mathrm{exact}}(b))/T_{\mathrm{exact}}(b)$, where $T_{\mathrm{exact}}$ in this case denotes the pairwise kSZ temperature signal using the exact cluster centers. We measure $B(b)$ for 1000 bootstrap realizations of the cluster sample, and in each realization we compute $\tilde{B}\equiv\langle B(b) \rangle_b$, i.e.\ the mean bias over all separation bins, in that realization. From the distribution of $\tilde{B}$ over all realizations we compute the mean bias, which we denote as $\bar{B}$, along with its standard deviation. 

In the top panel of Figure\,\ref{fig:misc} we show the pairwise kSZ signal as a function of comoving separation, for the different mis-centering models. As can be seen in that figure, mis-centering decreases the amplitude of the signal in all separation bins. In the bottom panel of that figure we show the mean bias $\bar{B}$ as a function of $\sigma_{\mathrm{J}}$, where the blue line indicates the value found in the Saro model. As expected, the absolute value of the bias increases for increasing $\sigma_{\mathrm{J}}$, reaching $-6.1\%$ in the case of $\sigma_{\mathrm{J}}=0.3\,\mpc$, and $-11.3\%$ for $\sigma_{\mathrm{J}}=0.5\,\mpc$. In the Saro model we find a bias of $-5.1$\%, which corresponds roughly to the Johnston model with $\sigma_J=0.3\mpc$.

Our results indicate that mis-centering does not have a significant impact on the SNR of the pairwise kSZ signal, as it will reduce the recovered signal by only $\lesssim$5\% in realistic models, and by up to $\sim 10\%$ in a pessimistic scenario ($\sigma_{\mathrm{J}}=0.5\,\mpc$.). The overall suppression of the amplitude in the signal due to mis-centering can however be a serious problem when fitting the data to a theoretical model because of the bias thus introduced. We note at this point that both mis-centering and astrophysical effects such as star-formation and feedback decrease the amplitude of the pairwise signal over the whole comoving separation range. Thus, if we want to use the pairwise kSZ signal for constraining gas models or cosmological parameters, prior knowledge about the mis-centering distribution is necessary.

\subsubsection{Scatter in mass}
\begin{figure}
\centering
\includegraphics[scale=0.4]{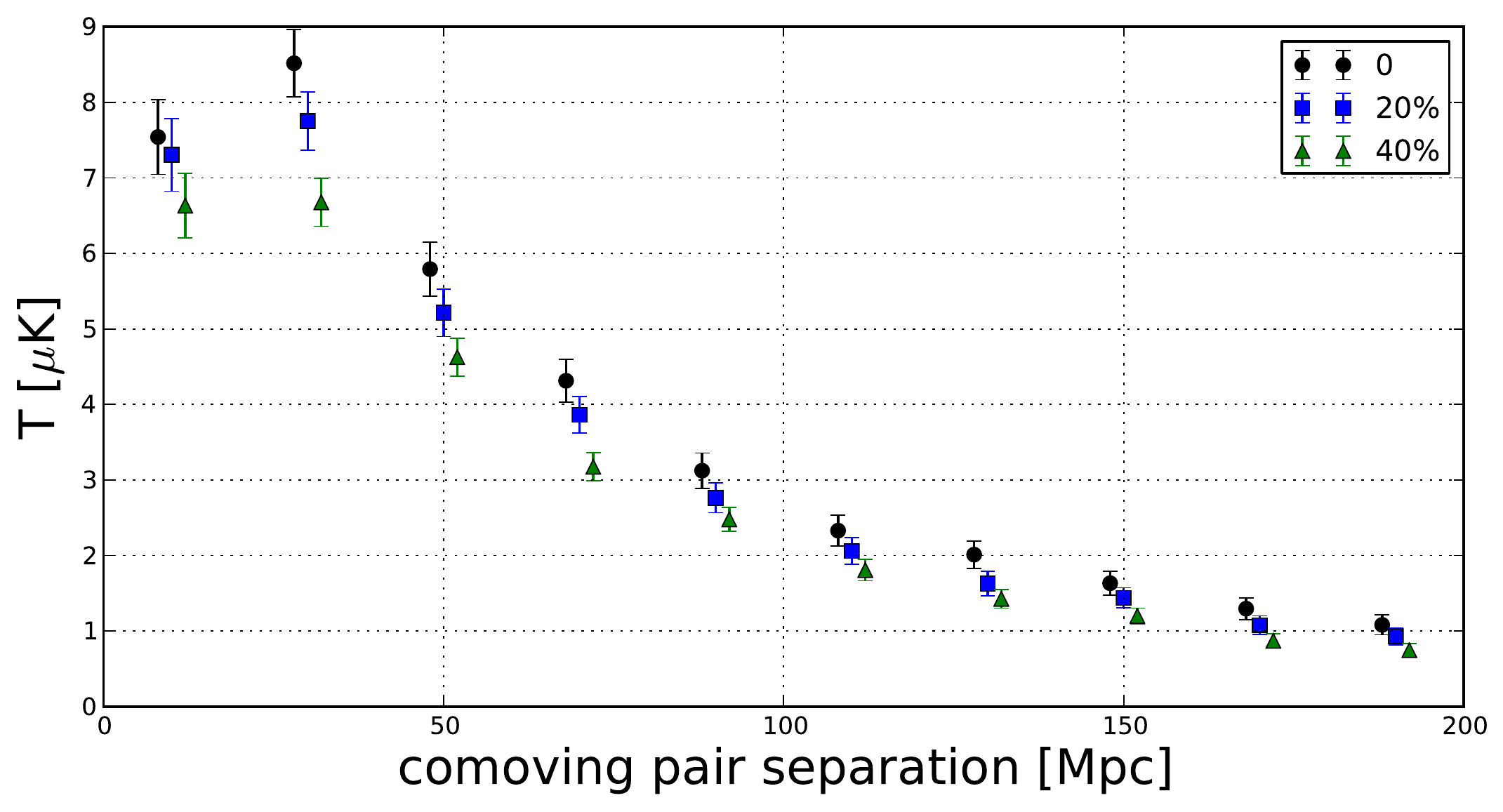}
\caption{\label{fig:mscatter} Pairwise kSZ signal as a function of comoving separation for different cluster samples, without noise components. The black points show the result for our baseline-model, $M_{200} \in (10^{14}, 3\cdot10^{14})\msol$. The blue and green points correspond to samples selected using $\tilde{M}_{200}$, i.e.,\ $M_{200}$ with an artificially introduced scatter of 20\% (blue) and 40\% (green). The amplitude of the signal decreases because the scatter in mass---in combination with the steepness of the mass-function---introduces far more low-mass clusters than high-mass clusters into the sample.}
\end{figure}

We have so far studied the pairwise kSZ signal for a cluster sample with a well-defined mass-cut, given in terms of $M_{200}$ (in our baseline-model: $M_{200}\in (10^{14}, 3\cdot10^{14})\msol$). In observations however, the cluster mass is generally unknown, and one therefore has to work with a proxy for the mass, e.g.,\ the richness of the cluster, which is roughly proportional to the number of member galaxies in the cluster (for details, see e.g.,\ \citealt{Rykoff:2011xi}). Given a richness-mass relation, the richness cut placed on the sample can be translated into a mass cut. There is however considerable scatter in the mass for a fixed richness, which---in combination with the steep mass-function---will preferably add low-mass clusters into the sample. Because low-mass clusters have a smaller kSZ amplitude, this introduces a bias into the sample that is similar to the well-known Eddington bias \citep{Eddington:1913,Eddington:1940}. Here, we wish to quantify that bias, as well as the change in the SNR introduced by it.

We assign to each cluster a new mass $\tilde{M}_{200}$, which we define by the relation $\ln \tilde{M}_{200}\equiv \ln M_{200} + \Delta$, where $\Delta$ is drawn from a Gaussian distribution with zero mean and standard deviation $\sigma_{\ln M}$. We consider two different values for $\sigma_{\ln M}$ here, 0.2 and 0.4, which correspond roughly to the range of expected values for the scatter in the richness-derived mass \citep{Rykoff:2011xi}. With $\sigma_{\ln M}=0.2$, we find that the size of the cluster sample increases from 11,700 (in our baseline sample) to 12,600, which corresponds to an increase of around 7\%. For $\sigma_{\ln M}=0.4$ the sample size increases to 15,100, i.e.,\ a 30\% increase. We plot the pairwise kSZ signal for different values of $\sigma_{\ln M}$ in Figure\,\ref{fig:mscatter}, using kSZ Model\,III without any additional noise components. As can be seen in this figure, the amplitude of the pairwise kSZ signal decreases for increasing $\sigma_{\ln M}$. As discussed above, the explanation for this effect is the combination of the scatter in mass with the steepness of the mass-function. We find that the amplitude decreases by $\sim$10-20\%, for $\sigma_{\ln M}=0.2$ and 0.4, respectively. Because of the increase in sample size, a slight increase in the SNR is expected; we find that the SNR increases from 8.3 in our baseline-model to 8.4 with $\sigma_{\ln M}=0.2$, and 8.9 with $\sigma_{\ln M}=0.4$. We note however that in real data the SNR might actually decrease with increasing scatter due to a lower purity of the cluster sample in this case.

\subsection{Pairwise kSZ signal with the compensated top-hat filter}

\begin{figure}
\centering
\includegraphics[scale=0.4]{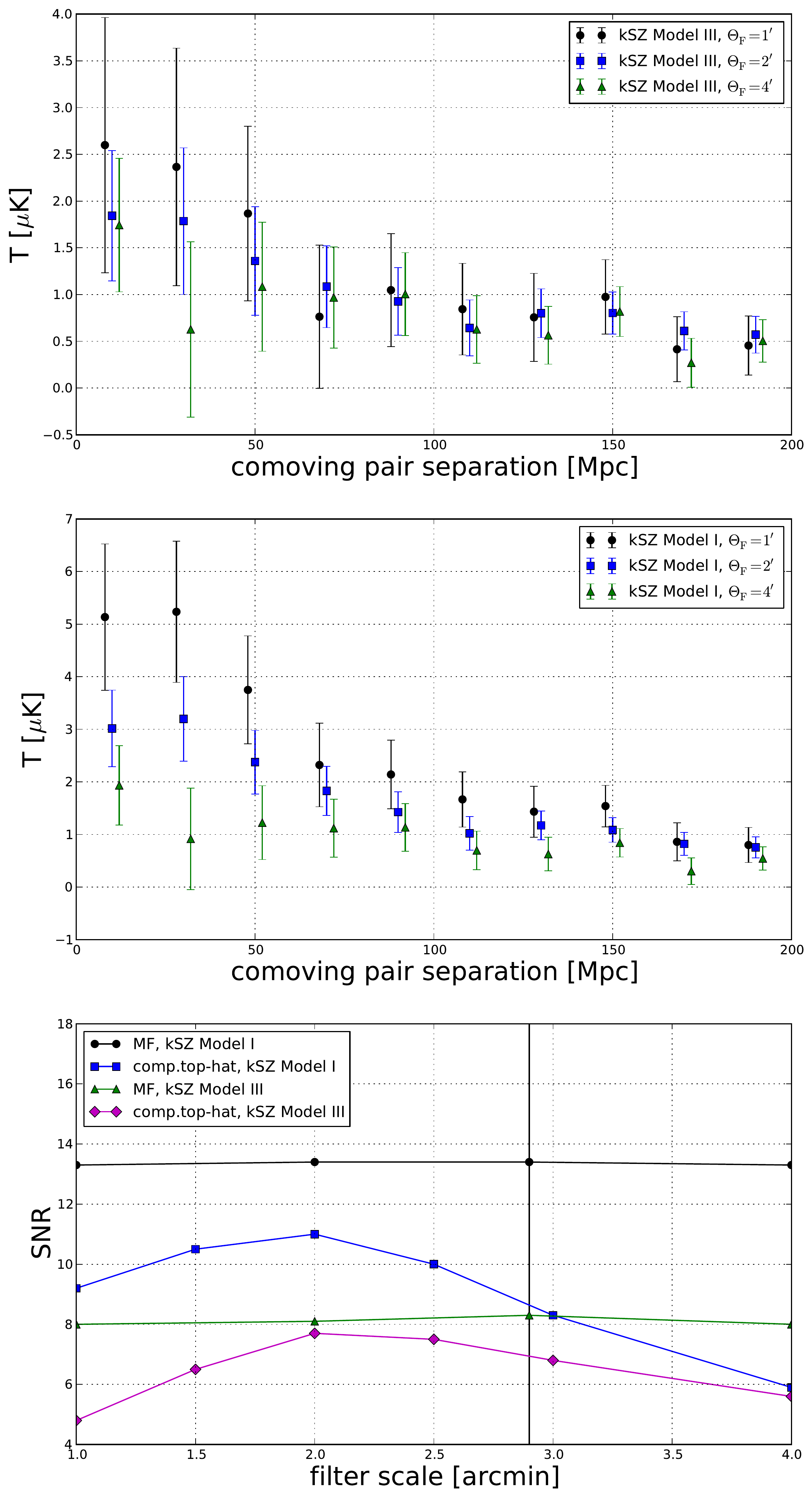}
\caption{\label{fig:compth} Pairwise kSZ signal as a function of comoving pair separation, using the compensated top-hat filter with different filter scales $\theta_{\rm F}$ for Model\,III (top panel)  and Model\,I (middle panel). Bottom panel: SNR as a function of $\theta_{\rm F}$. In both Model\,I and Model\,III, the SNR peaks at 2\,arcmin, which is close to 2.9\,arcmin---the average $\theta_{200}$ of the cluster sample (indicated by the black vertical line). For comparison, we show the SNR values found with the matched filter for Model\,III and Model\,I, as a function of the matched-filter parameter $\theta_{200}$. The SNR for the matched filter is essentially flat.}
\end{figure}

So far we have considered the results for the pairwise estimator using a matched filter. The other filter we test here is the compensated top-hat, which was used in the recent analysis of \cite{Ade:2015lza}. The authors chose this filtering technique---which does not make assumptions about the exact shape of the gas profile---to guard against potential biases given  uncertainties regarding the properties of the intra-cluster medium. Here, we ask how the performance of the compensated top-hat filter compares to the performance of the matched filter, which is tuned to the shape of the profile and the noise present in the data. We include both Model\,I and Model\,III in these tests. While we consider Model\,III our most accurate model, Model\,I (where ``gas traces mass'') is useful as it includes the full range of halo shapes and substructures that are produced in the $N$-body simulation. We note that for the tests involving Model\,III our results rely on the gas prescription from the Shaw model; different results might be found if a different gas prescription is used.

We vary the scale of the compensated top-hat filter from 1\,arcmin to 4\,arcmin. We find the best performance at a filter scale of 2\,arcmin, peaking at a significance of 7.5$\sigma$ for kSZ Model\,III and 11$\sigma$ for Model\,I (see bottom panel in Figure\,\ref{fig:compth}). For comparison, with the matched filter we found 8.3$\sigma$ for Model\,III and 13$\sigma$ for Model\,I. In both cases, the compensated top-hat filter performs slightly worse than the matched filter. We also varied the filter scale $\theta_{200}$ for the matched filter, and found that the SNR is essentially flat (see bottom panel in Figure\,\ref{fig:compth}), which can be explained by the fact that, for the clusters considered here, the beam-convolved projected NFW profiles with different $\theta_{200}$ parameters look very similar.

The average $\theta_{200}$ of the cluster sample is around 2.9\,arcmin and, as most of the gas is enclosed within that radius, the location of the signal-to-noise peak is thus expected around that scale. The exact position of the peak, however, depends on the gas physics. When comparing our results to experimental data or to simulations with a different gas prescription, a different peak scale might therefore be found. In fact, \cite{Ade:2015lza} report (weak) evidence for the pairwise kSZ signal using a compensated top-hat filter with a filter scale of up to three times $\theta_{200}$, contrary to expectations from simulations. The authors interpreted these findings as evidence of unbound gas significantly removed from the cluster center. As demonstrated in Figure\,\ref{fig:compth}, with our models (both Model\,I and Model\,III) we find a decrease of the SNR with increasing filter scale.

\subsection{Forecast for Next Generation Experiments}

\begin{table*}
\begin{center}
\begin{tabular}{|c|c|c|c|c|c|c|c|c|}
\hline
data scenario & overlap (sq.deg.) & $\sigma_{\mathrm{Instr}}$ (\muk-arcmin) & $\sigma_z/(1+z)$ & SNR Mod.\,III & SNR Mod.\,III* & SNR Mod.\,I & SNR Mod.\,I*\\ \hline \hline
SPT-SZ $\times$ DES     & 2500  & 18  & 0.01    & 4.4 & 6.2  & 7    & 9.8  \\ \hline
SPT-3G $\times$ DES     & 2500  & 2.5 & 0.01    & 6.6 & 14.7 & 9.5  & 22   \\ \hline
SPT-3G $\times$ SPHEREx & 2500  & 2.5 & 0.003** & 7   & 17   & 11.5 & 23.5 \\ \hline
Adv.ACTPol $\times$ DESI& 14000 & 8   & 0       &  20 & 27   & 36   & 57   \\ \hline
\end{tabular}
\end{center}
\caption{\label{table_forecasts}}
\begin{tablenotes}
\item Forecasts for the detection significance of the pairwise kSZ signal for current and future experiments, using kSZ Model\,I and Model\,III. For both models we consider a hypothetical case where the foregrounds and the tSZ component have been completely removed (indicated by a star). In practice, some of these components can be removed by combining frequency channels, in exchange for higher detector noise. In all scenarios we assume that all clusters in the range $M_{200} \in (10^{14}, 3\cdot10^{14})\msol$ and $z \in (0.2,1.0)$ are found in the survey. **For the scenario of SPT-3G $\times$ SPHEREx we assume that redshifts of clusters up to $z<0.45$ are measured with $\sigma_z/(1+z) = 0.003$, while the remaining redshifts are taken from DES data, with $\sigma_z/(1+z) = 0.01$.
\end{tablenotes}
\end{table*}

\begin{figure}
\centering
\includegraphics[scale=0.4]{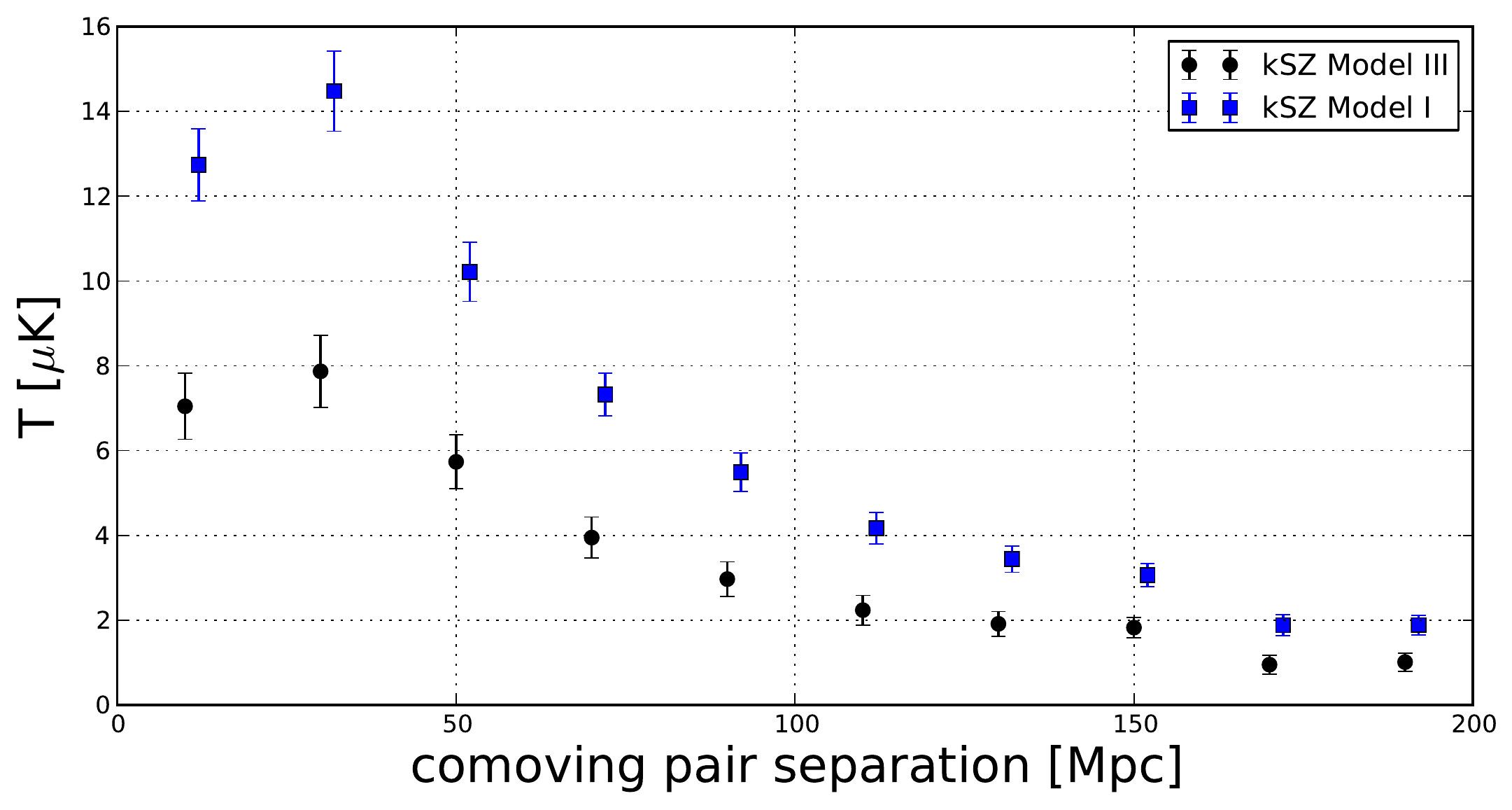}
\caption{\label{fig:s3} Pairwise kSZ signal in a data scenario that corresponds roughly to the combination of CMB data from Advanced ACTPol with galaxy clusters detected with DESI. We assume here a single-frequency measurement using the 150\,GHz channel with a detector noise of 8\,$\mu$K-arcmin, although in practice one could improve the significance by combining frequency channels. A high-significance detection is expected: we find a SNR of $\sim$20 for kSZ Model\,III, and $\sim$34 for Model\,I.}
\end{figure}

Let us finally estimate the detection significance of the pairwise kSZ signal from future optical and CMB experiments. For all of the scenarios discussed below we adopt the baseline cluster sample ($10^{14}\msol<M_{200}<3\cdot~10^{14}\msol$ and $0.2<z<1.0$); an overview of the considered scenarios is presented in Table\,\ref{table_forecasts}. As a reference, recall that our  analysis combining 150 GHz SPT-SZ data (noise 18\,$\mu$K-arcmin) with the DES cluster sample  (redshift errors of $\sigma_z/(1+z)=0.01$) found an SNR of $4.4-7$, when considering Model\,III and I, respectively.

We first consider the combination of DES data with that from SPT-3G, the next generation camera to be installed on the South Pole Telescope \citep{benson14}. 
SPT-3G will, like SPT-SZ, survey 2500 sq.deg.\ and have complete overlap with DES. It will have significantly lower instrument noise than SPT-SZ, with noise levels of [4, 2.5, 4]\,\muk-arcmin at [95, 150, 220]\,GHz, respectively. 
Considering 150 GHz alone, we find an SNR of 6.6-9.5 in this data scenario, corresponding to Model\,III and Model\,I, respectively. If we assume an idealized scenario where the 95 and 220 GHz data can be used to remove both foregrounds and the tSZ component without significant increase in instrumental noise, the SNR will increase up to $14.7-22$. For comparison, \citet{Keisler:2012eg} find $\mathrm{SNR}\sim18$ by optimally combining the 150 and 220 GHz data as a function of multipole to minimize the total power due to instrumental noise and emissive sources, and their result is within the boundaries stated above.

As discussed above, photometric errors greatly reduce the SNR of kSZ detections. One way to reduce these redshift errors on the DES cluster sample would be to incorporate data from SPHEREx \citep{Dore:2014cca}, a proposed all-sky spectroscopic survey satellite that would have complete overlap with SPT. At low redshifts, redshift errors at the level of $\sigma_z/(1+z) = 0.003$ are possible, and this improvement in the redshift measurements will lead to a higher detection significance. We assume that the redshift measurements of all DES clusters in the redshift range $z<0.45$ (corresponding to 24\% of the total cluster sample) can be improved in this way. In this scenario we estimate an SNR of $7-11.5$ when combining the improved DES sample with SPT-3G, and up to $17-23.5$ in the (hypothetical) case of perfectly cleaned CMB maps. These numbers will improve if methods to measure cluster redshifts with SPHEREx via averaging over galaxies prove to be successful, a topic we will consider in future investigations.

Finally, we model a data scenario that corresponds roughly to the combination of CMB data measured with Advanced ACTPol \citep{Calabrese:2014gwa, Henderson:2015nzj}, together  with the locations of galaxy clusters detected with the Dark Energy Spectroscopic Instrument (DESI, \citealt{Levi:2013gra}). We assume an overlap area of 14,000 sq.deg., and a CMB instrument noise level of 8~${\mu}$K-arcmin. We again consider the same mass cuts and redshift cuts as in our baseline model, and find $\sim$66,000 clusters within that volume. We find a pairwise kSZ signal with an SNR of $\sim$20 for this data scenario, using kSZ Model\,III (see Figure\,\ref{fig:s3}). We also simulate this scenario with kSZ Model\,I, and find an SNR of $\sim$34 in this case. Further cleaning of the maps can increase the SNR up to $27-57$, corresponding to Model\,III and I, respectively.

We have thus confirmed that high-significance detections of the pairwise kSZ signal are possible with future experiments. However, we need to be mindful that the cosmological implications of such measurements are limited by our lack of understanding of astrophysical processes, which can have significant impact on the measured amplitudes, as shown in this work.


\section{Discussion}

In this work, we have presented several different models for the kSZ signal, which are all implemented on the same $N$-body simulation and make different assumptions about the intra-cluster gas. Our first model, Model\,I, follows the simple assumption that baryons trace the dark matter and---given observational constraints---is likely an over-estimate of the true kSZ signal. We improve the realism in Model\,II which uses a gas prescription for each cluster based upon the model of \cite{Shaw:2010mn}, which is matched to observations. Finally, Model\,III, which is our most realistic model, combines the cluster component from Model\,II with the filament component from Model\,I. We have used these models in order to study the pairwise kSZ signal generated by clusters of galaxies. In particular, we confirm the results from \cite{Keisler:2012eg} that a high-significance detection of the pairwise kSZ signal is expected from high-resolution CMB data from SPT in conjunction with galaxy clusters detected with the complete 5-year DES survey. For our baseline-model, $M_{200} \in (10^{14}, 3\cdot10^{14})\msol$ and $z \in (0.2,1.0)$, we estimate a signal-to-noise ratio of 8-13, depending on the model, where the lower bound corresponds to kSZ Model\,III, and the upper bound to kSZ Model\,I. The SNR might be further improved by decreasing the mass threshold and thus increasing the size of the cluster sample. In the presence of photometric redshift errors the SNR is reduced to $\sim 4-7$ (depending on the model), using $\sigma_z=0.01$, and even further with larger values of $\sigma_z$.

One of our key results is that star formation and feedback modify the gas profile such that the amplitude of the pairwise kSZ signal is reduced by $\sim50$\% (see Figure\,\ref{fig:comp_models}), compared to Model~I. We have also shown that star formation and feedback reduce the kSZ power at small scales (see Figure\,\ref{fig:spectra}). Another important insight is that the pairwise kSZ signal with a matched filter on the cluster scale is mainly sensitive to the cluster-component of the kSZ, not the filamentary component. This becomes clear when comparing the results in Figure\,\ref{fig:comp_models} from Model\,II and Model\,III, which are indistinguishable. This is not surprising as the cluster component and the filament component of the kSZ are only correlated at large scales, and not on small scales (see Figure\,\ref{fig:spectra}).

Further, we have quantified the bias introduced in the amplitude of the pairwise kSZ signal due to systematic errors. We have found that photometric redshift errors introduce a bias that almost completely removes the pairwise signal at the lowest separation bins. Mis-centering on the other hand reduces the signal at all separation bins by up to 10\%. Another potential systematic effect is the scatter in the richness-derived mass. We have shown that this scatter, in combination with the steep mass function, introduces preferably low-mass clusters into the sample, decreasing the amplitude by $\sim 10-20\%$, for a scatter of $\sigma_{\ln M} = 0.2$ and 0.4, respectively. Our results stress that a good understanding of redshift errors, mis-centering, and the scatter in the richness-derived mass are needed for interpreting the amplitude of the pairwise kSZ signal as measured with real data.

We have also compared the performance of the pairwise estimator for two different filtering strategies, the matched filter and the compensated top-hat filter. We have found that the compensated top-hat filter at a filter-scale of 2~arcmin gives an SNR of $\sim 7.5-11$, depending on the kSZ model, which is slightly lower than the SNR of $8.3-13$ we found for the matched filter. We have demonstrated that the SNR peaks at a filter scale slightly below the average $\theta_{200}$ of the sample.

In this work, we have neglected the contribution from halo substructure and non-sphericity to the tSZ power spectrum. \citet{2012ApJ...758...75B} argue that these effects add 10-20\% to the tSZ power spectrum at high multipoles ($\ell>2000$), where most of the additional power is due to the halo substructure. The non-sphericity can add scatter to the $Y-M$ relation, as pointed out in \citet{White:2002wp}. Regarding the pairwise kSZ signal, we have demonstrated in this work that substructure and non-sphericity have no significant impact on the amplitude of the measured signal in an SPT-like survey; this can be seen by comparing the pairwise signal from Model\,I, which includes these effects, and Model\,II-NSF, which does not -- the results from these two models are consistent (see Figure~\ref{fig:comp_models}). It has also been shown in \citet{Nagai:2002nw} that velocity substructure adds only a scatter (of 50-100\,km/s) to the reconstructed peculiar cluster velocity, not a bias.

From a modeling perspective, we need to make significant progress in our understanding of the intra-cluster medium. The results presented here rely on the Shaw model for the gas profile, however, different results are to be expected for different astrophysical models and assumptions as the pairwise kSZ amplitude is strongly dependent on effects such as star formation and feedback. In particular, the kSZ amplitude would be larger by a factor of $\sim$2 in a model without star formation and feedback (see Figure~\ref{fig:comp_models}). Another important requirement for improved modeling of the pairwise kSZ signal is a better theoretical understanding (or modeling) of the BCG position distribution, which can help us to better understand and treat mis-centering.

Let us finally discuss some theoretical implications of our results. The pairwise kSZ signal probes the matter-velocity correlation function, which in Fourier space corresponds to the matter-velocity power spectrum $P_{v\delta}$. Using the continuity equation, $\mathbf{k}\cdot\mathbf{v}=-aHf\delta$, we have $P_{v\delta}=aHfP_{\delta\delta}/k$. The amplitude of the pairwise kSZ signal is thus proportional to the product $f\sigma_8^2$. However, as shown in this work, the astrophysical and systematic uncertainties in this amplitude are large. For this reason we conclude that it will be difficult to place constraints on these cosmological parameters from the pairwise kSZ signal alone. However, as these uncertainties affect only the overall amplitude of the signal (with the exception of redshift errors), it might be possible to use the measured shape of the pairwise kSZ signal as a function of separation in order to constrain non-standard cosmological models such as models that include modified gravity, dynamical dark energy, or non-zero neutrino masses \citep{Keisler:2012eg,Mueller:2014dba,Mueller:2014nsa}. The question then is whether the modified shape of the velocity cross-correlation function can still be measured in the presence of realistic noise and systematic uncertainties. We will address this question in future work, using $N$-body simulations of structure formation in cosmological models that go beyond standard $\Lambda$CDM~\citep{Heitmann:2015xma}.

\section*{Acknowledgments}
We acknowledge use of the software packages Healpix~\citep{2005ApJ...622..759G} and CAMB (http://camb.info/). We thank Suman Bhattacharya for helpful discussions and for contributing some of his code. Members of the HACC team---Nicholas Frontiere, Vitali Morozov, and Adrian Pope---made significant contributions to the simulation effort. We thank Martin White for helpful comments on a first draft of this work, and Olivier Dor\'e for discussions regarding SPHEREx. We further thank Tom Crawford, Ryan Keisler, Kyle Story, Bjoern Soergel, and Tommaso Giannantonio for helpful discussions related to this work. We thank the Referee for helpful comments on an earlier version of this work. This collaboration was initiated (2014 Summer Program) at the Aspen Center for Physics, which is supported by National Science Foundation grant PHY-1066293. Argonne National Laboratory's work was supported under the U.S. Department of Energy contract DE-AC02-06CH11357. This research used resources of the ALCF, which is supported by DOE/SC under contract DE-AC02-06CH11357 and resources of the National Energy Research Scientific Computing Center, a DOE Office of Science User Facility supported by the Office of Science of the U.S. Department of Energy under Contract DE-AC02-05CH11231. GH acknowledges funding from the National Sciences and Engineering Research Council of Canada and the Canadian Institute for Advanced Research.

\section*{Appendix}
\subsubsection*{A1: Matched filter normalization}
\begin{figure}
\centering
\includegraphics[scale=0.4]{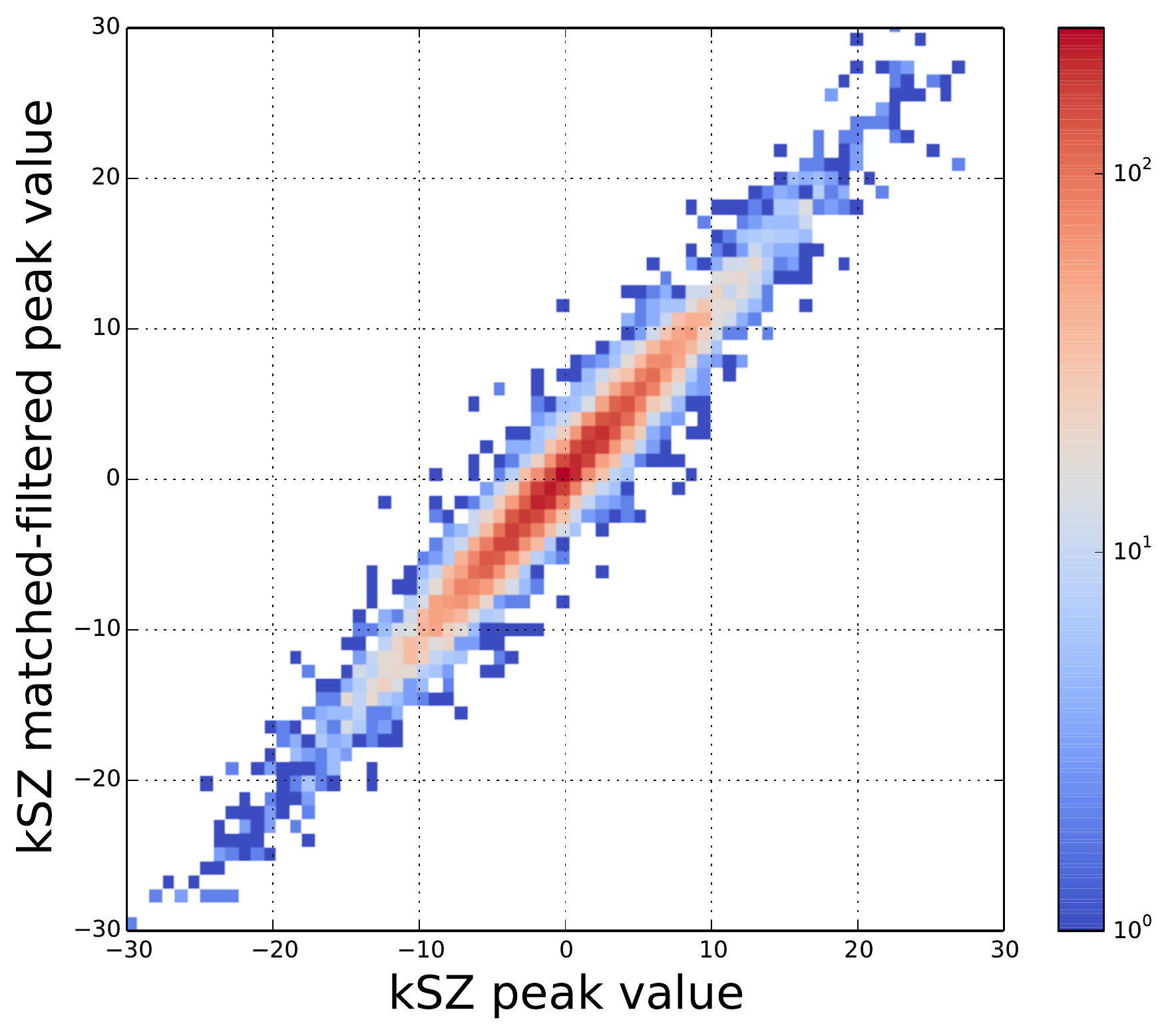}
\caption{\label{fig:mfnorm} 2D-histogram of temperature values at cluster locations in the matched-filtered kSZ (Model\,III) map (y-axis) vs the not matched-filtered, beam-convolved, kSZ map (x-axis). The distribution indicates that the matched filter correctly recovers the peak amplitude.}
\end{figure}

As explained in Section\,3.1.1, we normalize the beam-convolved matched-filter profile to unity at $\theta=0$. This normalization choice guarantees that the peak of the matched-filtered kSZ map at a cluster location is an unbiased estimate of the the observed (beam-convolved) kSZ peak. We test this normalization by plotting the matched-filtered peak values from the kSZ Model\,III map against the not matched-filtered, beam-convolved, peak values in Figure\,\ref{fig:mfnorm}: as expected, there is no indication of a bias in the matched-filtered peak values.

\subsubsection*{A2: Data products}
In this Appendix we provide some additional information concerning the data products created in this analysis, some of which we make publicly available. The halo-lightcone generated from all halos with more than 1000 particles up to $z=1$ takes up less than 1\,GB of disk-space, while the particle-lightcone is significantly larger (22\,TB). Note that this is only the size of a full-sky lightcone up to $z=1$. In principle, if so required, we could extend the lightcone up to higher redshifts, out to $z=10$, however, in the last case, the data thus created would require around 500\,TB of disk-space.

The maps corresponding to kSZ Model\,I, II, and III, the tSZ map, and the cluster catalog are publicly available at \texttt{http://www.hep.anl.gov/cosmology/ksz.html}. The maps are in Healpix ring-ordering with $\mathrm{Nside}=8192$, and have units of $\mu$K, relative to the average CMB temperature. These maps have a file-size of 3~GB each. The catalog is derived from the halo lightcone and has the format of a FITS file with the following FITS keys (all quantities are float, except \texttt{HALO\_ID}, which is a long int):

\bim 
\item \texttt{RA} - right ascension in degrees
\item \texttt{DEC} - declination in degrees
\item \texttt{M200RED} - $M_{200}$ (defined with respect to the critical density) in units of $\msol$
\item \texttt{M200} - $M_{200}$ in units of $\Msol$
\item \texttt{M500RED} - $M_{500}$ in units of $\msol$ (translated from \texttt{M200RED}, using the concentration \texttt{C200} (see below)
\item \texttt{M500} - $M_{500}$ in units of $\Msol$
\item \texttt{REDSHIFT} - redshift of the halo
\item \texttt{HALO\_ID} - identification number of the halo
\item \texttt{VLOS} - comoving peculiar line-of-sight velocity of the halo, in km/s
\item \texttt{C200} - concentration assigned to the halo based on $M_{200}$, following the concentration-mass relation from \citet{2013ApJ...766...32B}
\eim 

\bibliography{bib.bib}

\end{document}